\documentclass[onecolumn,pra,superscriptaddress,notitlepage,a4paper,nofootinbib,accepted=2021-07-30]{quantumarticle}

\pdfoutput=1

\usepackage[numbers,sort&compress]{natbib}
\usepackage[dvipsnames,svgnames]{xcolor}
\usepackage{url}
\usepackage{hyperref}

  \usepackage{morefloats}
 \usepackage{graphicx}
 \usepackage{dcolumn}%
 \usepackage{bm,bbm}%
 \usepackage{braket}
\usepackage{amsthm,amsmath,amssymb,amsbsy}
\usepackage{enumerate}
\usepackage{subcaption,caption}
\usepackage{booktabs,multirow}
\usepackage{mathtools,mathrsfs,bbding}
\usepackage[percent]{overpic}
\usepackage{microtype}

\usepackage[T1]{fontenc}
\usepackage[utf8]{inputenc}

\usepackage{array,adjustbox,afterpage}

\newtheoremstyle{red}{}{}{\itshape}{}{\color{red!80!black}\bfseries}{.}{ }{}

\definecolor{darkred}{rgb}{0.57,0,0.12}

\let\LL\L
\let\ll\l

\usepackage{cleveref}
\usepackage[most,breakable]{tcolorbox}

\let\nc\newcommand

\newcommand{\br}[1]{{\color{blue!35!red}#1}}

\let\br\relax

\DeclareMathOperator{\Tr}{Tr}

\DeclareMathOperator{\NN}{NN}

\renewcommand{\H}{\mathcal{H}}

\newcommand{\norm}[2]{\left\lVert#1\right\rVert_{\,#2}}
\newcommand{\proj}[1]{\ket{#1}\!\bra{#1}}

\newcommand{\snorm}[2]{\big\lVert#1\big\rVert_{\,#2}}

\newcommand{\T}{\mathcal{T}}

\newcommand{\A}{\mathcal{A}}
\nc{\Ka}{{\mathcal{K}_{\alpha}}}
\nc{\Kad}{{\mathcal{K}_{\alpha}\hspace{-0.8ex}\raisebox{0.2ex}{\*}}}
\nc{\Kadp}{{\mathcal{K}_{\alpha+}\hspace{-1.8ex}\raisebox{0.2ex}{\*}\hspace{.7ex}}}
\nc{\Kadd}{{\mathcal{K}_{\alpha}\hspace{-0.8ex}\raisebox{0.2ex}{\*\*}}}
\nc{\Kaddd}{{\mathcal{K}_{\alpha}\hspace{-0.8ex}\raisebox{0.2ex}{\*\*\*}}}

\newcommand{\C}{\mathcal{C}}
\newcommand{\G}{\mathcal{G}}
\newcommand{\U}{\mathcal{U}}

\newcommand{\F}{\mathcal{F}}
\newcommand{\D}{\mathcal{D}}
\renewcommand{\L}{\mathcal{L}}

\renewcommand{\P}{\mathcal{P}}
\newcommand{\M}{\mathcal{M}}
\newcommand{\N}{\mathcal{N}}

\newcommand{\Z}{\mathcal{Z}}

\newcommand{\CPTP}{{\mathrm{CPTP}}}
\newcommand{\SPA}{{\mathrm{SPA}}}

\newcommand{\CP}{{\mathrm{CP}}}
\newcommand{\CPTN}{{\mathrm{CPTNI}}}

\let\CPTNI\CPTN

\renewcommand{\*}{\textup{*}}
\newcommand{\<}{\left\langle}
\renewcommand{\>}{\right\rangle}

\renewcommand{\bar}{\;\rule{0pt}{9.5pt}\right|\;}

\newcommand{\lset}{\left\{\left.}
\newcommand{\rset}{\right\}}

\newcommand{\RR}{\mathbb{R}}

\newcommand{\HH}{\mathbb{H}}

\newcommand{\DD}{\mathbb{D}}
\renewcommand{\NN}{\mathbb{N}}

\newcommand{\ve}{\varepsilon}

\newcommand{\id}{\mathbbm{1}}
\newcommand{\idc}{\mathrm{id}}

\renewenvironment{boxed}[1]%
  {\expandafter\ifstrequal\expandafter{#1}{red}{\begin{tcolorbox}[colback=red!3,colframe=red!15]}{\begin{tcolorbox}[colback=white,colframe=gray!10,breakable,enhanced]}}%
  {\end{tcolorbox}}

\let\textsc\uppercase

{\begin{boxed}{orange}\vspace{-\baselineskip}\begin{equation}}%
{\end{equation}\end{boxed}}

\theoremstyle{definition}

\newtheorem{theorem}{Theorem}

\newtheorem{proposition}[theorem]{Proposition}
\newtheorem{corollary}[theorem]{Corollary}

\newtheorem{lemma}[theorem]{Lemma}
\newtheorem*{remark}{Remark}

\theoremstyle{red}

\let\oldproofname\proofname
\renewcommand{\proofname}{\rm\bf{\oldproofname}}

  \nc{\MIO}{{\text{\rm MIO}}}
\nc{\DIO}{{\text{\rm DIO}}}
\nc{\SIO}{{\text{\rm SIO}}}
\nc{\IO}{{\text{\rm IO}}}

\nc{\lsetr}{\left\{\,}
\nc{\rsetr}{\right.\right\}}
\nc{\barr}{\;\rule{0pt}{9.5pt}\left|\;}
\nc{\ketbra}[2]{\ket{#1}\!\bra{#2}}

\nc{\wt}{\widetilde}

\nc{\dia}{\!\!\Diamond}
\nc{\bdia}{\!\!\blacklozenge}

\let\bnorm\bdia

\newcommand{\dm}[1]{\ketbra{#1}{#1}}
\newcommand{\bal}{\begin{equation}\begin{aligned}}
\newcommand{\eal}{\end{aligned}\end{equation}}

\begin{document}

 \title{%
 Operational applications of the diamond norm and related measures in quantifying the non-physicality of quantum maps
 }%

\author{Bartosz Regula}
\email{bartosz.regula@gmail.com}
\affiliation{Nanyang Quantum Hub, School of Physical and Mathematical Sciences, Nanyang Technological University, 637371, Singapore}
\orcid{0000-0001-7225-071X}

\author{Ryuji Takagi}
\email{ryuji.takagi@ntu.edu.sg}
\affiliation{Nanyang Quantum Hub, School of Physical and Mathematical Sciences, Nanyang Technological University, 637371, Singapore}
\orcid{0000-0003-3837-8159}

\author{Mile Gu}
\email{mgu@quantumcomplexity.org}
\affiliation{Nanyang Quantum Hub, School of Physical and Mathematical Sciences, Nanyang Technological University, 637371, Singapore}
\affiliation{Complexity Institute, Nanyang Technological University, 637371, Singapore}
\affiliation{Centre for Quantum Technologies, National University of Singapore, 3 Science Drive 2, 117543, Singapore}

\begin{abstract}
Although quantum channels underlie the dynamics of quantum states, maps which are not physical channels --- that is, not completely positive --- can often be encountered in settings such as entanglement detection, non-Markovian quantum dynamics, or error mitigation. We introduce an operational approach to the quantitative study of the non-physicality of linear maps based on different ways to approximate a given linear map with quantum channels. 
Our first measure directly quantifies the cost of simulating a given map using physically implementable quantum channels, shifting the difficulty in simulating unphysical dynamics onto the task of simulating linear combinations of quantum states. Our second measure benchmarks the quantitative advantages that a non-completely-positive map can provide in discrimination-based quantum games.
Notably, we show that for any trace-preserving map, the quantities both reduce to a fundamental distance measure: the diamond norm, thus endowing this norm with new operational meanings in the characterisation of linear maps. We discuss applications of our results to structural physical approximations of positive maps, quantification of non-Markovianity, and bounding the cost of error mitigation. 
\end{abstract}
 \maketitle

\section{Introduction}

It is one of the fundamental properties of quantum mechanics that the evolution of quantum states is described by linear maps which are completely positive and trace preserving (CPTP), stemming from the unitary dynamics enforced on a larger Hilbert space~\cite{nielsen_2011}. However, in several different settings of practical importance, various applications of quantum dynamics which are not CPTP can be encountered. This motivates the study of such transformations, and in particular a precise understanding of how they can be compared with and approximated by physical quantum channels.

One important application of non-CPTP maps is in entanglement detection, where positive but not completely positive maps can serve as entanglement witnesses~\cite{horodecki_1996-1}. A bipartite state $\rho$ is entangled if and only if there exists a positive map $\Phi$ such that $\idc \otimes \Phi(\rho)$ is no longer a positive operator, and therefore such a map can reveal the correlations of $\rho$. This approach has constituted one of the most important ways of detecting entanglement~\cite{guhne_2009,horodecki_2009}, but its experimental implementation encounters an obstacle: how to realise the action of an unphysical linear map in practice? This question prompted the introduction of structural physical approximations (SPA) of non-CPTP maps~\cite{horodecki_2003-4}, which aim to enable the physical evaluation of general maps by designing suitable approximations in terms of quantum channels and using them to infer properties of the original map~\cite{horodecki_2002-1,korbicz_2008,bae_2017}.

Another setting in which non-CPTP maps are encountered is that of non-Markovian quantum dynamics or, generally, in the reduced dynamics of correlated systems. Specifically, when an open quantum system shares some initial correlations with its environment, the evolution of the composite system-environment state can correspond to a non-CPTP map when looking only at the dynamics of the reduced state of the system~\cite{pechukas_1994,shaji_2005,rodriguez-rosario_2008,carteret_2008}. Although the physical interpretation of this is a matter of debate and alternative ways to understand such dynamics have been proposed~\cite{alicki_1995,modi_2012-1,schmid_2019}, it can nevertheless be useful to study such non-CPTP evolutions directly to gain an understanding of reduced dynamics of open quantum systems.

Even broader types of unphysical quantum dynamics can be found in the areas of quantum error correction and error mitigation~\cite{shor_1996,temme_2017,li_2017}. This is because, in a broad sense, both of these settings are concerned with the following problem: if an unknown system has undergone a noisy evolution as $\rho \mapsto \Theta(\rho)$, how can we reconstruct the original state as closely as possible, that is, how to implement a map $\Phi$ such that $\Phi \circ \Theta (\rho) \approx \rho$? Such inverse operations typically cease to be valid quantum channels, and so it becomes necessary to devise approaches to implement them in practice with the use of physical operations.

In this work, we introduce a general quantitative framework for the characterisation of such unphysical maps by approximating them with quantum channels. We then explicitly give the considered measures operational meaning by connecting them with the performance of practical tasks, including the cost of simulating a given map with quantum channels. Notably, we show that all of the considered measures reduce to the same quantity when the given linear map is trace preserving: they all equal the diamond norm~\cite{kitaev_1997,watrous_2004}, a fundamental computational tool that serves as a measure of quantum channel distance and finds many uses in the practical characterisation of quantum processes~\cite{watrous_2018}. This endows the diamond norm with new meanings in the operational tasks that we consider, and furthermore allows a number of new connections to be established. On the one hand, many known results in the quantification of the diamond norm can be carried over to the setting of our work, and on the other hand, we can use our characterisation to provide new insight into the computation and applications of the diamond norm.

Our approach is based on the notion of robustness measures~\cite{vidal_1999} --- inspired by recent applications of such quantities in the study of general resource theories of channels~\cite{diaz_2018-2,takagi_2019,liu_2019-1,gour_2019-1,uola_2020,yuan_2020,takagi_2020,takagi_2020-2,regula_2020-2,regula_2020-3,jiang_2020}, we use them to quantify the amount of noise needed to turn a given map into a quantum channel. Such measures allow for several different generalisations to the setting of linear maps, motivating us to study and compare these definitions. The robustness-based approaches can be understood as different ways of designing optimal decompositions of linear maps in terms of quantum channels, and so they generalise the standard structural physical approximations~\cite{horodecki_2003-4}. We express the measures as semidefinite programs and establish various relations and bounds between them.

We apply our first measure in the task of simulating the action of an unphysical map with valid channels, accomplished by allowing the use of ancillary systems which can consists of linear combinations of quantum states. \br{Such an approach allows us to reduce the problem of simulating the dynamics of quantum systems to the much simpler case of simulating the use of a non-positive Hermitian operator. Assessing the difficulty of this procedure then reduces to quantifying how much the given operator deviates from being a valid quantum state, and --- employing the trace norm as a natural quantifier of such `non-quantumness' --- }
we show that the optimal cost of simulating a non-CPTP map in this way is given exactly by the value of the robustness measure.

Furthermore, answering the question of whether any unphysical map can provide measurable operational advantages over quantum channels, we show this to be the case in the setting of discrimination-based quantum games, establishing our second robustness measure as the exact quantifier of this advantage.

Our results also generalise and shed light on the very recent findings of Ref.~\cite{jiang_2020}, which considered a similar framework for approximating trace-preserving maps using a robustness- and quasiprobability-based approach. In particular, we show that the measure considered in \cite{jiang_2020} is actually an alternative expression for the diamond norm of a map, rather than a new quantity.

The paper is structured as follows. In Sec.~\ref{sec:robustness_intro}, we introduce the notions of robustness measures and show how they can be applied to non-CPTP linear maps. We establish precise connections with the diamond norm in Sec.~\ref{sec:diamond}. We then proceed to show that the robustness measures --- and hence the diamond norm --- play a crucial role in quantifying the cost of simulating linear maps (Sec.~\ref{sec:simulation}) as well as in understanding the advantages a non-CPTP map could provide in input-output quantum games (Sec.~\ref{sec:games}). We proceed to establish a number of bounds for the measures in Sec.~\ref{sec:bounds}. Finally, we discuss the applications of our approach, comparisons with other methods, and explicitly show how the measures can be evaluated for some representative examples in Sec.~\ref{sec:applications}.

\section{Robustness of non-CP maps}\label{sec:robustness_intro}

Let $A$ and $B$ denote two finite-dimensional quantum systems of dimension $d_A$ and $d_B$, respectively. We will use $\mathbb{L}(A)$ to denote the set of all linear operators, $\HH(A)$ to denote the set of all Hermitian operators, and $\DD(A)$ to denote all density operators acting on the Hilbert space of system $A$. We use $\< X, Y \> = \Tr(X^\dagger Y)$ for the Hilbert-Schmidt inner product.

Among all linear maps from $\mathbb{L}(A)$ to $\mathbb{L}(B)$, we will be primarily concerned with Hermiticity-preserving maps $\H(A,B)$, which are defined as maps such that $\Phi(X) \in \HH(B) \; \forall X \in \HH(A)$. A map is called positive if $\Phi(X) \geq 0 \; \forall X \geq 0$ (w.r.t.\ the positive semidefinite cone), completely positive (CP) if $\idc_{A} \otimes \Phi$ is positive, trace preserving if $\Tr \Phi(X) = \Tr X \; \forall X$, and trace non-increasing if $\Tr \Phi(X) \leq \Tr X \; \forall X$. To each map $\Phi \in \H(A,B)$ we will associate the Choi operator $J_\Phi = (\idc_A \otimes \Phi)[\proj{\Omega}] \in \HH(A \otimes B)$ where $\ket{\Omega} = \sum_{i} \ket{ii}$. Importantly, a map is Hermiticity-preserving iff $J_\Phi = J_\Phi^\dagger$, CP iff $J_{\Phi} \geq 0$, and trace preserving iff $\Tr_B J_\Phi = \id_A$ (see e.g.\ \cite{watrous_2018}). Let $\CPTN(A,B)$ denote the set of completely positive and trace--non-increasing maps in $\H(A,B)$, and analogously $\CPTP(A,B)$ the set of completely positive and trace-preserving maps. For simplicity of notation, we will often simply write $\CPTP$ for $\CPTP(A,B)$ (and analogously for other sets) when the spaces in consideration are not relevant.

In order to quantify how much a given map deviates from the set of CPTP maps, we will employ the concept of \textit{robustness measures}~\cite{vidal_1999}. It will be insightful to first review how such measures are defined for quantum states. Given a \br{convex} set of interest $\F \subseteq \DD$, commonly chosen to be the set of free states in a given resource theory, one asks: how much noise from a set $\N \subseteq \DD$ has to be added to a state $\rho$ in order to make it a free state? This has the intuitive interpretation of measuring how robust the resources contained in the state $\rho$ are with respect to noise from the set $\N$. Specifically, we write
\begin{equation}\begin{aligned}
  r_\N(\rho) \coloneqq \min \lsetr \lambda \barr \frac{\rho + \lambda \omega}{1+\lambda} \br{\eqqcolon \sigma} \in \F,\; \omega \in \N \rsetr.
\end{aligned}\end{equation}
The most common choices of the noise set $\N$ are: $\N = \DD$, in which case we obtain the so-called \textit{generalised robustness} equivalently given by
\begin{equation}\begin{aligned}
  r_\DD(\rho) = \min \lset \lambda \bar \rho \leq (1+\lambda) \sigma,\; \sigma \in \F \rset,
\end{aligned}\end{equation}
and the choice $\N = \F$, which corresponds to the \textit{standard robustness} $r_\F$. The latter quantity is directly related to the so-called base norm $\norm{\rho}{\F}$ of the set $\F$, which can be alternatively understood as an optimisation of quasiprobability distributions over the set $\F$:
\begin{equation}\begin{aligned}
  2 r_\F(\rho) + 1  =& \norm{\rho}{\F}\\
  \coloneqq&  \min \lset \lambda_+ + \lambda_- \bar \rho + \lambda_- \sigma_- = \lambda_+ \sigma_+,\; \sigma_{\pm} \in \F \rset\\
  =& \min \lset \sum_i |\lambda_i| \bar \rho = \sum_i \lambda_i \sigma_i,\; \sigma_{i} \in \F \rset\\
\end{aligned}\end{equation}
where the third line is a simple consequence of the convexity of $\F$. The definitions straightforwardly extend to unnormalised operators $X$: \br{defining
\begin{equation}\begin{aligned}
r_\N(X) \coloneqq \min \lsetr \lambda \barr \frac{X + \lambda \omega}{\Tr X +\lambda} \eqqcolon \sigma \in \F,\; \omega \in \N \rsetr,
\end{aligned}\end{equation}
}%
it is important to notice that the trace of $X$ will come into play, and the base norm will equal $\norm{X}{\F} = 2 r_\F(X) + \Tr X$.

The case of interest to us will be where the set of free states $\F$ contains \textit{all} physical quantum states, $\F = \DD$, in which case the different notions of the robustness are equal and one has
\begin{equation}\begin{aligned}
  2 r_\DD(X) + \Tr X = \norm{X}{1},
\end{aligned}\end{equation}
that is, the base norm is precisely the trace norm (Schatten 1-norm) $\norm{\cdot}{1}$.

\paragraph{Robustness of linear maps.} A generalisation of these concepts to the case of linear maps can be done in several different ways. Firstly, one has to note that it does not suffice to consider trace-preserving maps \br{in the definitions of this measures. This follows since any linear combination of CPTP maps necessarily satisfies that $\Tr_B J_\Phi \propto \id$, which means that $\Tr \Phi(\rho)$ takes the same value for any input state $\rho$. Therefore, any Hermiticity-preserving map whose reduced Choi matrix is not proportional to the identity operator cannot be represented as $\lambda_+ \Lambda_+ - \lambda_- \Lambda_-$ for CPTP $\Lambda_\pm$.} To circumvent this, we will employ the set of completely positive and trace--non-increasing maps, which can be understood as probabilistic implementations of quantum channels. Importantly, robustness-based definitions which were all equal in the case of states might not be equal any more. We therefore need to explicitly consider three different types of the robustness w.r.t.\ the sets $\CPTP$ or $\CPTN$:
\begin{align}
  R(\Phi) \coloneqq& \min \lsetr \lambda \barr \frac{\Phi + \lambda \Lambda}{1+\lambda} \in \CPTN,\; \Lambda \in \CPTN \rsetr,\label{eq:rob1}\\
  R'(\Phi) \coloneqq& \min \lsetr \lambda \barr J_\Phi \leq (1+\lambda) J_\Lambda,\; \Lambda \in \CPTN \rsetr\label{eq:rob2}\\
  =& \min \lsetr \lambda \barr J_\Phi \leq (1+\lambda) J_\Lambda,\; \Lambda \in \CPTP \rsetr,\nonumber\\
  R''(\Phi) \coloneqq& \min \lset \lambda \bar \Phi + \lambda \Lambda \in \CP,\; \Lambda \in \CPTN \rset\label{eq:rob3}\\
  =& \min \lset \lambda \bar \Phi + \lambda \Lambda \in \CP,\; \Lambda \in \CPTP \rset,\nonumber
\end{align}
as well as a generalised notion of a base norm with respect to the set of completely positive and trace--non-increasing maps:
\begin{equation}\begin{aligned}
  \norm{\Phi}{\bnorm} \coloneqq \min \lset \lambda_+ + \lambda_- \bar \Phi = \lambda_+ \Lambda_+ - \lambda_- \Lambda_-,\; \Lambda_{\pm} \in \CPTN \rset.
  \label{eq:CPTNI norm def}
\end{aligned}\end{equation}
In the expressions for $R'$ and $R''$, we made use of the fact that one can, without loss of generality, restrict the optimisation to CPTP maps; this follows since for any $\Lambda \in \CPTN$ such that $\Tr_B J_\Lambda \leq \id_A$ we can define the map $\Lambda'$ by $J_{\Lambda'} =  J_\Lambda + \frac{C}{d_B} \otimes \id_{B}$ where $C = \id_A - \Tr_B J_\Lambda \geq 0$ which satisfies $\Lambda' \in \CPTP$ and achieves the same value of the objective function. We note that closely related definitions were recently also considered in Ref.~\cite{jiang_2020} for the case of trace-preserving maps.

All of the quantities above are well-defined and take a finite value for any map $\Phi \in \H(A,B)$, as we shall see explicitly by establishing general upper bounds in Sec.~\ref{sec:bounds}. \br{The robustness $R(\Phi)$ can be seen to be an upper bound for all other quantities: any feasible decomposition of $\Phi$ in Eq.~\eqref{eq:rob1} gives feasible solutions for Eq.~\eqref{eq:rob2}, \eqref{eq:rob3}, and for the base norm in Eq.~\eqref{eq:CPTNI norm def}.} It is a priori unclear whether one can find general conditions under which the inequalities between the different measures are tight. We shall shortly see that equality indeed holds for all trace-preserving linear maps.

All of the introduced quantities can be computed as semidefinite programs, which follows since the constraints for a map to be CPTNI (or CPTP) are linear matrix inequalities. This means that the measures can be evaluated efficiently (in the dimensions of the map) using numerical software. The equivalent dual forms of the problems, which can also provide some insight into the differences between the different definitions of the robustness measures, will be reported shortly in Sec.~\ref{sec:bounds}.


\section{Relation with the diamond norm}\label{sec:diamond}

For any Hermiticity-preserving map $\Phi$, the diamond norm (completely bounded trace norm) is defined as~\cite{kitaev_1997,watrous_2018}
\begin{equation}\begin{aligned}
  \norm{\Phi}{\dia} = \max_{\rho \in \DD(A \otimes A)} \norm{ \idc_A \otimes \Phi \, (\rho)}{1},
\end{aligned}\end{equation}
where, in a slight abuse of notation, we use $\DD(A \otimes A)$ to denote the states acting on a bipartite Hilbert space composed of the space $A$ and another space isomorphic thereto.

The diamond norm finds use as a fundamental measure of distance between quantum channels, mirroring the operational role of the trace distance in measuring distances between quantum states~\cite{kitaev_1997,Sacchi2005,gilchrist_2005,watrous_2018}. It is one of the most widely employed figures of merit in comparing quantum channels and benchmarking channel manipulation protocols. Its quantification and characterisation is therefore crucial to an effective understanding of the properties of quantum processes.
\br{Close connections between the diamond norm and the base norm in the space of quantum channels can be inferred already from the operational similarity that the diamond norm bears to the trace norm, the latter being the natural base norm in the space of quantum states. Here we aim to clarify the details of such connections and to explicitly relate the diamond norm with the robustness measures.
}

We will first introduce the following lemma, which establishes a useful formulation of the diamond norm for Hermiticity-preserving maps. The result is closely related to a more general approach for generalised quantum channels considered previously by Jenčová~\cite{jencova_2014}, \br{and can be alternatively deduced from Lem.~4 and Thm.~2 of \cite{jencova_2014}}.

\begin{boxed}{white}
\begin{lemma}\label{lemma:diamond_herm}
For any Hermiticity-preserving map $\Phi$, it holds that
\begin{align}\label{eq:diamond_tighter_ineq}
  \norm{\Phi}{\dia} &= \min \lset \mu \bar J_\Phi = M_+ - M_-,\; M_{\pm} \geq 0,\; \Tr_B (M_+ + M_-) \leq \mu \id_{A}  \rset\\
  &= \min \lset \mu \bar J_\Phi = M_+ - M_-,\; M_{\pm} \geq 0,\; \Tr_B (M_+ + M_-) = \mu \id_{A}  \rset.\label{eq:diamond_tighter}
\end{align}
\end{lemma}
\end{boxed}
\begin{proof}
Let $\norm{\Phi}{\dia}'$ denote the quantity in \eqref{eq:diamond_tighter}. We first notice that the constraint $\Tr_B (M_+ + M_-) = \mu \id_A$ can be relaxed to $\Tr_B (M_+ + M_-) \leq \mu \id_A$  without loss of generality. This follows since for any feasible $M_\pm$ s.t. $\Tr_{B} (M_+ + M_-) + C = \mu \id$ with $C\geq 0$, one can define feasible solutions $M'_{\pm} = M_\pm + \frac{C}{2d_B} \otimes \id_{B}$ which satisfy $\Tr_B (M'_+ + M'_-) = \mu \id_{A}$ and thus achieve the same optimal value. We thus have
\begin{equation}\begin{aligned}\label{eq:diamond_ch_sdp}
  \norm{\Phi}{\dia}' &= \min \lset \mu \bar J_\Phi = M_+ - M_-,\; M_{\pm} \geq 0,\; \Tr_B (M_+ + M_-) \leq \mu \id_{A}  \rset\\
  &= \min \lset \norm{\Tr_B (M_+ + M_-)}{\infty} \bar J_\Phi = M_+ - M_-,\; M_{\pm} \geq 0 \rset.
\end{aligned}\end{equation}
Taking the Lagrange dual of the above (see Appendix~\ref{app:duals}) gives
\begin{equation}\begin{aligned}\label{eq:norm2}
  \norm{\Phi}{\dia}' &= \max \lset \< J_\Phi, W \> \bar - \rho \otimes \id_B \leq W \leq  \rho \otimes \id_B,\; \rho \in \DD(A) \rset\\
  &=  \sup \lset \< J_\Phi, W \> \bar - \rho \otimes \id_B \leq W \leq  \rho \otimes \id_B,\; \rho \in \DD_{>0}(A) \rset\\
  &= \sup \lset \< W, \sqrt{\rho \otimes \id_B} \, J_\Phi\, \sqrt{\rho \otimes \id_B} \> \bar \rho \in \DD_{>0}(A),\, -\id_{A \otimes B} \leq W \leq \id_{A \otimes B} \rset\\
  &= \sup_{\rho \in \DD_{>0}(A)} \norm{\sqrt{\rho \otimes \id_B} \, J_\Phi\, \sqrt{\rho \otimes \id_B}}{1}\\
  &= \max_{\rho \in \DD(A)} \norm{\sqrt{\rho \otimes \id_B} \, J_\Phi\, \sqrt{\rho \otimes \id_B}}{1},
\end{aligned}\end{equation}
where in the second line, \br{by continuity, we restricted our attention to the set of full-rank states $\DD_{>0}(A)$ without loss of generality}, and in the third line we made the change of variables $W \mapsto \sqrt{\rho^{-1}\otimes \id_B} W \sqrt{\rho^{-1}\otimes \id_B}$. The fact that this equals the diamond norm of $\Phi$ can be deduced from the results of Ref.~\cite{watrous_2009} already; for completeness, we will show this explicitly. Recalling that $J_\Phi = (\idc_A \otimes \Phi)\proj{\Omega}$ with $\ket{\Omega}$ being the unnormalised maximally entangled state, and using the fact that $\Phi$ is only acting on one of the subsystems, we can write
\begin{equation}\begin{aligned}
   \norm{\Phi}{\dia}' &= \max_{\rho \in \DD(A)} \norm{ \sqrt{\rho \otimes \id_B} \, (\idc_A \otimes \Phi) \left[ \proj{\Omega} \right]\, \sqrt{\rho \otimes \id_B}}{1}\\
   &= \max_{\rho \in \DD(A)} \norm{ (\idc_A \otimes \Phi) \left[ \sqrt{\rho} \otimes \id_A \, \proj{\Omega}\, \sqrt{\rho} \otimes \id_A \right]}{1}\\
   &= \max_{\psi \in \DD(A \otimes A)} \norm{ \idc_A \otimes \Phi \,(\psi)}{1}\\
   &= \max_{\rho \in \DD(A \otimes A)} \norm{ \idc_A \otimes \Phi \,(\rho)}{1}\\
   &= \norm{\Phi}{\dia}
\end{aligned}\end{equation}
where we used that any pure state $\psi \in \DD(A \otimes A)$ can be written as $\left(\sqrt{\rho} \otimes \id_A\right)  \proj{\Omega} \left(\sqrt{\rho} \otimes \id_A\right)$ for a suitable choice of $\rho \in \DD(A)$, with $\ket{\psi}$ constituting the canonical purification of $\rho$. 
\end{proof}

\br{Compared with the semidefinite programs for the diamond norm of general linear maps originally derived in Refs.~\cite{watrous_2009,watrous_2013}, the form of the diamond norm presented in Lemma~\ref{lemma:diamond_herm} already constitutes a major simplification --- both at a conceptual level, allowing for a restatement of the problem in terms of optimising over decompositions of the form $J_\Phi = M_+ - M_-$, and computationally, as the number of optimisation variables is reduced.}

As an immediate consequence of the above result, we can use the characterisation of the diamond norm in Eq.~\eqref{eq:diamond_tighter_ineq} to construct valid feasible solutions for the base norm and robustness measures in Eqs.~\eqref{eq:rob1}--\eqref{eq:CPTNI norm def}, and vice versa.
\begin{boxed}{white}
\begin{corollary}\mbox{}\label{cor:dia_cptni_ineq}
For any Hermiticity-preserving map $\Phi$, it holds that
  \begin{equation}\begin{aligned}\label{eq:dia_cptni_ineq}
 2  \norm{\Phi}{\dia} &\geq \norm{\Phi}{\bnorm} \geq \norm{\Phi}{\dia},\\
 \norm{\Phi}{\dia} + 1 &\geq R'(\Phi) \geq \frac{1}{2}\left(\vphantom{\Big[}\norm{\Phi}{\dia} - 2 + \lambda_{\min}(\Tr_B J_\Phi)\right)\\
  \norm{\Phi}{\dia} &\geq R''(\Phi) \geq \frac{1}{2}\left(\vphantom{\Big[}\norm{\Phi}{\dia} - \lambda_{\max}(\Tr_B J_\Phi)\right),
\end{aligned}\end{equation}
\end{corollary}
where $\lambda_{\min}$ and $\lambda_{\max}$ denote, respectively, the smallest and the largest eigenvalues.
\end{boxed}
\br{%
\begin{proof}
Any decomposition for the diamond norm of the form $J_\Phi = M_+ - M_-$ with $\Tr_B (M_+ + M_-) \leq \mu \id_{A}$ satisfies $\Tr_B M_{\pm} \leq \mu \id_A$, which provides valid feasible solutions to the norm $\norm{\cdot}{\bnorm}$ and the robustness measures. On the other hand, any decomposition for $\norm{\cdot}{\bnorm}$ of the form $ \Phi = \lambda_+ \Lambda_+ - \lambda_- \Lambda_-$ with $\Lambda_{\pm} \in \CPTN$ gives a feasible decomposition for the diamond norm with $\Tr_B (\lambda_+ J_{\Lambda_+} + \lambda_- J_{\Lambda_-}) \leq (\lambda_+ + \lambda_-)\id_A$. Similarly, any decomposition for $R'$ satisfying $J_\Phi \leq (1+\lambda) J_\Lambda$ gives a feasible decomposition for $\norm{\cdot}{\dia}$ of the form $J_\Phi = (1+\lambda) J_\Lambda - M_-$ where $M_- \coloneqq (1+\lambda) J_\Lambda - J_\Phi$. Using that
\begin{equation}\begin{aligned}
\Tr_B\left[(1+\lambda) J_\Lambda + M_-\right] &= \Tr_B \left[ 2(1+\lambda) J_\Lambda - J_\Phi\right]\\
&\leq \left[2(1+\lambda) - \lambda_{\min}(\Tr_B J_\Phi)\right] \id_A,
\end{aligned}\end{equation}
we get the stated bound. The case of $R''$ follows analogously.
\end{proof}
}%

Equality between the different quantities can be shown for all trace-preserving maps, directly relating the diamond norm with our considered measures.

\begin{boxed}{white}
\begin{theorem}\label{thm:rob_diamond}
For any map $\Phi \in \H(A,B)$ which is trace preserving or, more generally, proportional to a trace-preserving map in the sense that $\Tr_B J_\Phi \propto \id$, it holds that
\begin{equation}\begin{aligned}\label{eq:dia_tp}
  \norm{\Phi}{\dia} =  \norm{\Phi}{\bnorm} = \min \lset \mu_+ + \mu_- \bar J_\Phi = \mu_+ J_{\Lambda_+} - \mu_- J_{\Lambda_-},\; \Lambda_{\pm} \in \CPTP ,\; \mu_{\pm} \in \RR_+ \rset.
\end{aligned}\end{equation}
\br{For trace-preserving maps $\Phi$, it additionally holds that}
\begin{equation}\begin{aligned}
  R(\Phi) = R'(\Phi) = R''(\Phi) = \frac{\norm{\Phi}{\bnorm}-1}{2} = \frac{\norm{\Phi}{\dia}-1}{2}.
\end{aligned}\end{equation}
\end{theorem}
\end{boxed}
\begin{proof}
From the fact that $\Tr_B J_\Phi = t \id_A$ for some $t \in \RR$, it is easy to see that every decomposition of the form $J_\Phi = M_+ - M_-,\; M_{\pm} \geq 0,\; \Tr_B (M_+ + M_-) = \mu \id_{A}$ as in Lemma~\ref{lemma:diamond_herm} has to satisfy
\begin{equation}\begin{aligned}
  \Tr_B M_+ = \frac{\mu+t}{2} \id_A,\quad \Tr_B M_- = \frac{\mu-t}{2} \id_A.
\end{aligned}\end{equation}
This implies that we can equivalently write
\begin{equation}\begin{aligned}
  \norm{\Phi}{\dia} &= \min \lset \mu_+ + \mu_- \bar J_\Phi = \mu_+ M_+ - \mu_- M_-,\; M_{\pm} \geq 0,\; \Tr_B M_+ = \Tr_B M_- = \id_{A}  \rset
\end{aligned}\end{equation}
which is precisely Eq.~\eqref{eq:dia_tp}. Notice then that any such decomposition gives a valid feasible solution for $\norm{\Phi}{\bnorm}$, together with Cor.~\ref{cor:dia_cptni_ineq} yielding equality between the two norms.

When $\Phi$ is trace preserving ($t=1$), we can write
\begin{equation}\begin{aligned}\label{eq:dia_tp_eq}
  \norm{\Phi}{\dia} &= \min \lset \mu_+ + \mu_- \bar J_\Phi = \mu_+ J_{\Lambda_+} - \mu_- J_{\Lambda_-},\; \Lambda_{\pm} \in \CPTP \rset\\
  &= \min \lset 2 \mu_+ - 1 \bar J_\Phi = \mu_+ J_{\Lambda_+} - \mu_- J_{\Lambda_-},\; \Lambda_{\pm} \in \CPTP \rset\\
  &= \min \lset 2 \mu_- + 1 \bar J_\Phi = \mu_+ J_{\Lambda_+} - \mu_- J_{\Lambda_-},\; \Lambda_{\pm} \in \CPTP \rset.
\end{aligned}\end{equation}
The equality $\norm{\Phi}{\dia} = 2 R'(\Phi) + 1 = 2 R''(\Phi) + 1$ then follows: on the one hand, any decomposition of the form in Eq.~\eqref{eq:dia_tp_eq} gives a feasible decomposition for $R'$ and $R''$ in Eqs.~\eqref{eq:rob2}--\eqref{eq:rob3}, and on the other hand, any decomposition for the robustness measures is necessarily of the form in Eq.~\eqref{eq:dia_tp_eq}. 
Equality with the robustness $R(\Phi)$ follows by noting again that any feasible decomposition in Eq.~\eqref{eq:dia_tp_eq} gives a feasible decomposition for $R(\Phi)$, and on the other hand using the relation $R(\Phi) \geq R'(\Phi)$ which holds by definition.
\end{proof}

\begin{remark}The expression in Eq.~\eqref{eq:dia_tp} is valid also in the case of trace-annihilating maps ($\Tr_B J_\Phi = 0$), and thus the case of computing the distance $\norm{\Lambda - \Lambda'}{\dia}$ between two quantum channels. A simplified expression for this problem appeared previously in~\cite{watrous_2009} and was explicitly expressed as a robustness-type measure in~\cite{gour_2019-1}.
\end{remark}

We note that the quantity $\norm{\cdot}{\bnorm}$, applied to trace-preserving maps, was recently considered in Refs.~\cite{jiang_2020} and~\cite{piveteau_2021}. It was not noticed in these works that this is simply the diamond norm, and hence many results shown in \cite{jiang_2020} (e.g.\ the multiplicativity with respect to tensor product, unitary invariance, bounds with trace norm $\norm{J_\Phi}{1}$, monotonicity under the action of superchannels, and some explicit expressions) follow directly from known properties of the diamond norm~\cite{watrous_2004,watrous_2009,michel_2018,nechita_2018}.

We will later see that this equivalence does not extend to maps which are not trace preserving (or proportional thereto), and indeed we can have $\norm{\Phi}{\bnorm} = 2 \norm{\Phi}{\dia}$ in the extreme case.


\section{Quantifying simulation cost}\label{sec:simulation}

Since the quantum dynamics which can be realised in practice are restricted to completely positive maps, a relevant question then becomes: how can one simulate the action of a non-CPTP map on a quantum state when only CPTP maps are available to us?

A similar question was recently asked in Ref.~\cite{jiang_2020}, where the authors applied quasiprobability sampling methods~\cite{pashayan_2015,temme_2017,takagi_2020-2} to the desired operation $\Phi$. We take a different approach here and instead allow for the use of an ancillary system $X$, which can be an affine combination of quantum states, in order to simulate the action of the map $\Phi$ as a CPTP map $\Lambda$ acting jointly on the input quantum state and the ancilla $X$. The ``non-physicality'' of the given map $\Phi$ is then pushed into the system $X$, allowing for the overall transformation $\Lambda$ to be a valid quantum channel.

The motivation for this approach is that the task of simulating the action of the non-CPTP map $\Phi$ is effectively replaced with the simulation of a unit-trace Hermitian operator $X$, which could be significantly easier to realise in practice, especially since we will see that the dimension of the ancilla can be taken to be arbitrarily small. Standard quasiprobability-based approaches such as the ones employed in \cite{temme_2017,takagi_2020-2,jiang_2020} aim to estimate the expectation value $\Tr[\Phi(\rho)A]$, where $\Phi$ is a non-CPTP map and $A$ an observable, by decomposing the given map as $\Phi = \lambda_i \Lambda_i$ with $\lambda_i \in \RR$ and $\Lambda_i \in \CPTP$ (or CPTNI). The expectation value $\Tr[\Phi(\rho)A]$ is then estimated by evaluating $\Tr[\Lambda_i(\rho)A]$ and appropriately sampling from the output distributions with probabilities determined by the coefficients $\lambda_i$~\cite{pashayan_2015,temme_2017}. In practice, this means that we have to repeatedly realise each operation $\Lambda_i$, which requires the implementation of a different quantum circuit for each operation. Consider, on the other hand, a situation in which the dynamics is fixed as some map $\Lambda \in \CPTN$, and we only need to vary the input states. This can be achieved by writing $\Phi(\cdot) = \Lambda(\cdot \otimes X)$, where we can write any Hermitian operator in a quasiprobability representation as $X = \sum_i \mu_i \rho_i$. The task of sampling from the output distribution is then reduced to feeding in the different states $\rho_i$ into the circuit which realises the fixed operation $\Lambda$, thus greatly simplifying the implementation.

As mentioned in Sec.~\ref{sec:robustness_intro}, a natural quantifier of how much a given operator $X \in \HH(A)$ with $\Tr(X) = 1$ deviates from the set of all quantum states is the trace norm $\norm{X}{1}$. Indeed, this quantity can be given an explicit interpretation in terms of the optimal cost of a quasiprobability-based \br{estimation of the expectation value of $X$}~\cite{pashayan_2015}. We then define the simulation cost of a map as the minimal amount of such ``non-physicality'' of $X$ needed to simulate the action of the map:
\begin{equation}\begin{aligned}
  S(\Phi) \coloneqq \min \lset \norm{X}{1} \bar \Lambda (\cdot \otimes X ) = \Phi(\cdot), \;\Tr(X) = 1,\; \Lambda \in \CPTN \rset.
\end{aligned}\end{equation}

We then have the following.
\begin{boxed}{white}
\begin{theorem}\label{thm:cost}
For any map $\Phi \in \H(A,B)$, it holds that
\begin{equation}\begin{aligned}
  S(\Phi) = 2 R(\Phi) + 1.
\end{aligned}\end{equation}
In the case of a trace-preserving $\Phi$, we have in particular that
\begin{equation}\begin{aligned}
  S(\Phi) = \norm{\Phi}{\dia}
\end{aligned}\end{equation}
and an optimal $\Lambda$ for the simulation can be chosen to satisfy $\Lambda \in \CPTP$.
\end{theorem}
\end{boxed}
\begin{proof}
Let $\Lambda_{\pm} \in \CPTN(A,B)$ be maps that achieve an optimal decomposition for $\Phi$ such that $\Phi=(1+R(\Phi))\Lambda_+ - R(\Phi)\Lambda_-$. 
Now, consider a non-positive Hermitian operator $X= \mu_+ \omega_+ - \mu_- \omega_- \in \HH(A')$ where $\omega_\pm$ are orthogonal quantum states and $\Tr[X]= \mu_+ - \mu_- =1$. We do not impose any additional conditions on the size of the ancillary system $A'$, meaning that its Hilbert space can be chosen to be an arbitrary space of dimension at least 2. Defining the projector onto the positive part of $X$ as $P_+$, we then consider the map defined by the action on a basis $\ketbra{i_1}{j_1}\otimes \ketbra{i_2}{j_2}\in \mathbb{L}(A \otimes A')$ as follows:
\begin{equation}\begin{aligned}
 \Lambda(\ketbra{i_1}{j_1}\otimes\ketbra{i_2}{j_2})&:=\Tr\left[\frac{P_+}{\mu_+}\ketbra{i_2}{j_2}\right]\Phi(\ketbra{i_1}{j_1})+\Tr\left[\left(\id-\frac{P_+}{\mu_+}\right)\ketbra{i_2}{j_2}\right]\Lambda_-(\ketbra{i_1}{j_1})\\
 &= (1+R(\Phi))\Tr\left[\frac{P_+}{\mu_+}\ketbra{i_2}{j_2}\right]\Lambda_+(\ketbra{i_1}{j_1})\\
 &\quad +\Tr\left[\left(\id-(1+R(\Phi))\frac{P_+}{\mu_+}\right)\ketbra{i_2}{j_2}\right]\Lambda_-(\ketbra{i_1}{j_1})
 \label{eq:direct construction}
\end{aligned}\end{equation}

It is easy to check that $\Lambda(\rho\otimes X)=\Phi(\rho)$.
Now, we will show that as long as the condition
\begin{eqnarray}
(1+R(\Phi))\Tr\left[\frac{P_+}{\mu_+}\rho\right]\leq 1\; \forall \rho
\label{eq:direct positive}
\end{eqnarray}
is satisfied, then $\Lambda$ is also CPTNI.
This can be seen by observing first that \eqref{eq:direct positive} gives
\begin{eqnarray} 
0\leq \tilde{P} \coloneqq (1+R(\Phi))\frac{P_+}{\mu_+}\leq \id,
\end{eqnarray}
which implies that $\tilde{P}$ is a valid POVM element.
Note that we can rewrite \eqref{eq:direct construction} as 
\begin{eqnarray}
 \Lambda&=& \Lambda_+\otimes T_{\tilde{P}} +\Lambda_-\otimes T_{\id-\tilde{P}} \\
 &=& (\id\otimes T_{\tilde{P}})(\Lambda_+\otimes\id) +(\id\otimes T_{\id-\tilde{P}})(\Lambda_-\otimes\id) 
\end{eqnarray}
where $T_{P}(\cdot):=\Tr\left[P\cdot \right]$.
Since $\Lambda_+$, $\Lambda_-$, and $T_{\tilde{P}}$, $T_{\id-\tilde{P}}$ are all completely positive, $\Lambda$ is also completely positive. 
Since \eqref{eq:direct positive} is always satisfied when
\begin{eqnarray}
\mu_+ \geq 1+R(\Phi),
\end{eqnarray}
an operator $X$ with $\norm{X}{1} = \mu_+ + \mu_- = 1+2 R(\Phi)$ achieves the desired implementation.

The converse part can be proven by extending an argument in Ref.~\cite{diaz_2018-2} to our setting. Suppose a non-quantum resource $X=\mu_+ \omega_+ - \mu_- \omega_-$ and CPTNI map $\Lambda$ realise the simulation of $\Phi$, i.e. $\Phi(\cdot) = \Lambda\left(\cdot\otimes X\right)$.
Also, define $\Lambda_+(\cdot) \coloneqq\Lambda(\cdot\otimes \omega_+)$.
Then, by linearity of $\Lambda$, we get
\begin{eqnarray}
 \Lambda_+(\cdot)&=&\frac{1}{\mu_+}\Lambda(\cdot\otimes X)+\frac{1}{\mu_+}\Lambda(\cdot\otimes \mu_- \omega_-)\\
 &=& \frac{1}{\mu_+}\Phi(\cdot)+\frac{\mu_-}{\mu_+}\Lambda(\cdot\otimes \omega_-).
\end{eqnarray}
Since $\Lambda_+$ and $\Lambda_- \coloneqq \Lambda(\cdot\otimes \omega_-)$ are CPTNI maps, this is a valid linear decomposition of $\Phi$ into two CPTNI maps, providing an upper bound for its robustness as $R(\Phi)\leq \mu_- = \mu_+-1$. This gives the desired lower bound for the simulation cost as $\mu_+ + \mu_- \geq 1 + 2 R(\Phi)$. 
\end{proof}

An interesting quantitative equivalence emerges between our approach and the method of Ref.~\cite{jiang_2020}. In that work, the authors showed that the minimal overhead required to employ quasiprobability-based simulation techniques~\cite{pashayan_2015,temme_2017} to estimate $\Tr[\Phi(\rho)A]$ for a trace-preserving map $\Phi$ scales with the norm $\norm{\Phi}{\bnorm}$ (see also the discussion in Sec.~\ref{sec:examples_inverses}). Since we know from Thm.~\ref{thm:rob_diamond} that
\begin{equation}\begin{aligned}
  2R(\Phi) + 1 = \norm{\Phi}{\bnorm} = \norm{\Phi}{\dia}
\end{aligned}\end{equation}
holds for any trace-preserving map, the quantitative cost of the simulation scheme is actually the same as our method, despite the seemingly different approaches employed. In fact, our Thm.~\ref{thm:cost} shows that it is sufficient to consider decompositions of $\Phi$ as
\begin{equation}\begin{aligned}\label{eq:quasiprob_ancilla}
  \Phi(\cdot) = \mu_+ \Lambda(\cdot \otimes \omega_+) - \mu_- \Lambda(\cdot \otimes \omega_-)
\end{aligned}\end{equation}
where $\Lambda$ and $X = \mu_+ \omega_+ - \mu_- \omega_-$ are as constructed in our protocol. This means that, despite the significant practical simplification obtained by fixing the dynamics of the simulator as $\Lambda$ and optimising over the quasiprobability representations of $X$ instead, our simulation method does not sacrifice any performance, and the optimal sampling overhead cost of the more direct approach of \cite{jiang_2020} cannot be any better.  

We note that Theorem~\ref{thm:cost} gives a general way of reducing the task of simulating the action of a linear map $\Phi$ to simulating an affine combination of states in the form of the operator $X$. This could provide methods for the simulation of dynamics even beyond quasiprobability-based approaches like the one discussed above, although the specifics of this will depend on the given simulation method.

\paragraph{State injection and resource simulation.} The setting considered here is closely related to state injection methods which generalise quantum teleportation~\cite{bennett_1993} and find use e.g.\ in the resource theories of entanglement~\cite{berta_2013,pirandola_2017,wilde_2018,gour_2019,bauml_2019}, stabiliser-state quantum computation~\cite{gottesman_1999,seddon_2019}, and coherence~\cite{bendana_2017,diaz_2018-2}. In such tasks, a resourceful state $\phi$ (such as a maximally entangled singlet) is used to simulate the action of an arbitrary quantum channel $\Theta$ as $\Theta(\cdot) = \Gamma(\cdot \otimes \phi)$, where now $\Gamma$ is a free operation (such as a protocol consisting of local operations and classical communication only). In this sense, our result can be thought of as the cost of channel simulation in the resource theory of ``non-physicality'' beyond quantum mechanics, with the operator $X$ acting as a resource. There are many potential ways to interpret such a result: for instance, unit-trace Hermitian operators which are not necessarily positive semidefinite have found use as so-called pseudo-states in~\cite{geller_2014}, where they were used to study correlations beyond quantum mechanics, and as so-called pseudo-density matrices in~\cite{fitzsimons_2015}, where they were used to put spatial and temporal correlations on equal footing. Being able to use a Hermitian system $X$ could then be interpreted as having access to such extended sets of correlations. We leave a precise investigation of the connections between the operational setting employed here and resource theories of correlations for future work.

\paragraph{Amortised simulation.} A related setting that we can consider is that of \textit{amortised} simulation \cite{kaur_2017,diaz_2018-2}, in which the non-quantum resource $X$ is not consumed completely, but instead we can recover some of it in the form of another resource $Y$ which can be reused. Precisely, we define
\begin{equation}\begin{aligned}
  S_A(\Phi) = \min \lset \frac{\norm{X}{1}}{\norm{Y}{1}} \bar \Lambda (\cdot \otimes X ) = \Phi(\cdot) \otimes Y, \;\Tr(X) = \Tr(Y) = 1,\; \Lambda \in \CPTN \rset.
\end{aligned}\end{equation}
\br{Clearly, $S_A(\Phi) \leq S(\Phi)$ as we can just take $X$ to be optimal for $S$ and $Y$ to be the trivial system $1$. One could expect amortisation to lead to a strictly smaller cost of simulating a given map. However, we can show that this is not the case --- amortisation cannot improve the simulation cost of any trace-preserving map.}

\begin{boxed}{white}
\begin{corollary}\label{corr:cost_amo}
  For any trace-preserving map $\Phi \in \H(A,B)$, it holds that
\begin{equation}\begin{aligned}
  S_A(\Phi) = S(\Phi) = \norm{\Phi}{\dia}.
\end{aligned}\end{equation}
\end{corollary}
\end{boxed}
\begin{proof}
Let $\Lambda$ be the optimal map such that $\Lambda (\cdot \otimes X ) = \Phi(\cdot) \otimes Y$ with $S_A(\Phi) = \norm{X}{1}/\norm{Y}{1}$. \br{Noting that this can be alternatively understood as a simulation protocol for the trace-preserving map $\Phi(\cdot) \otimes Y$, Thm.~\ref{thm:cost} tells us that any such protocol satisfies}
\begin{equation}\begin{aligned}
  \norm{X}{1} &\geq S\left( \Phi(\cdot) \otimes Y \right)\\
  &= \norm{ \Phi(\cdot) \otimes Y }{\dia}\\
  &= \norm{ \Phi}{\dia} \norm{Y}{1}\\
  &= S (\Phi) \norm{Y}{1}
\end{aligned}\end{equation}
where we used the multiplicativity of the  diamond norm and the fact that $\norm{Y}{\dia}=\norm{Y}{1}$ where we treat $Y$ as a preparation channel with a trivial input space. From this we have that $S_A(\Phi) \geq S(\Phi)$, which concludes the proof.
\end{proof}


\section{Quantifying advantages in quantum games}\label{sec:games}

The study of general linear maps in a resource-theoretic setting motivates the question: is there a well-defined operational task in which having access to \emph{any} non-CPTP map could provide practical advantages over all quantum channels?

In order to give an instance of such a task, we consider the setting of input-output games, inspired by the work of Ref.~\cite{rosset_2018} and studied in the context of dynamical quantum resources in~\cite{uola_2020,yuan_2020}. The setting is as follows: Alice prepares a state chosen randomly from the ensemble $\{p_i, \sigma_i\}_i$ and sends the state through the map $\Phi \in \H(A,B)$ to Bob, who then measures with a POVM $\{M_j\}_j$. The players are then awarded a score based on a reward function characterised by the coefficients $\{w_{ij}\}_{i,j} \in \RR$, and their goal is to maximise the average payoff given by
\begin{equation}\begin{aligned}
  P(\Phi, \{p_i, \sigma_i\}, \{M_j\}, \{w_{ij}\}) = \sum_{i,j} w_{ij} p_i \< M_j, \Phi(\sigma_i) \>
\end{aligned}\end{equation}
by a suitable choice of the states and measurements. The tuple $\G = (\{p_i, \sigma_i\}, \{M_j\}, \{w_{ij}\})$ then defines the \textit{input-output game} $\G$.

We stress that, although the payoff $P(\Phi, \G)$ might lose its physical meaning as a discrimination task when $\Phi$ is an arbitrary linear map, already for a \textit{positive} trace-preserving map $\Phi$ we have that every output $\Phi(\sigma_i)$ is indeed a valid density matrix and thus the measurement at the output constitutes a well-defined state discrimination task.

We are then interested in quantifying the best possible advantage that a given map $\Phi$ could provide over CPTP maps. Such an optimisation is unbounded without any further constraints, so we will consider games for which any completely positive map $\Gamma$ achieves a non-negative payoff value --- this can always be ensured by suitably shifting the payoff function for a given game. We then have the following.
\begin{boxed}{white}
\begin{theorem}\label{thm:operational}
For any map $\Phi \in \H(A,B)$, it holds that
\begin{equation}\begin{aligned}\label{eq:operational_diam}
  \sup_{\G} \frac{P(\Phi, \G)}{\max_{\Lambda \in \CPTP} P(\Lambda, \G)} = R'(\Phi) + 1
\end{aligned}\end{equation}
where the maximisation is over all input-output games $\G$ such that $P(\Gamma,\G) \geq 0 \; \forall \Gamma \in \CP$. In the case of a trace-preserving $\Phi$, we have in particular that
\begin{equation}\begin{aligned}
  \sup_{\G} \frac{P(\Phi, \G)}{\max_{\Lambda \in \CPTP} P(\Lambda, \G)} =  \frac{\norm{\Phi}{\dia}+1}{2},
\end{aligned}\end{equation}
and it suffices to optimise over games such that $P(\Lambda,\G) \geq 0 \; \forall \Lambda \in \CPTP$.
\end{theorem}
\end{boxed}
\begin{proof}
Any $\Phi$ can be written as $\Phi = (1+R'(\Phi)) \Lambda - \Gamma$ where $\Lambda \in \CPTP$, $\Gamma \in \CP$. On the one hand, we then have for any $\G$ that
\begin{equation}\begin{aligned}
  P(\Phi, \G) &= (1+R'(\Phi)) \,P(\Lambda, \G) -P(\Gamma, \G)\\
  &\leq (1+R'(\Phi))\, P(\Lambda, \G)\\
  &\leq (1+R'(\Phi)) \max_{\Lambda \in \CPTP} P(\Lambda, \G)
\end{aligned}\end{equation}
where the first inequality follows since $P(\Gamma,\G) \geq 0 \; \forall \Gamma \in \CP$, which shows that the left-hand side of Eq.~\eqref{eq:operational_diam} is upper-bounded by the right-hand side. By Thm.~\ref{thm:rob_diamond}, in the case of a trace preserving map $\Phi$ we can equivalently write $\Phi = (1+R'(\Phi)) \Lambda_+ - R'(\Phi) \Lambda_-$ where $\Lambda_\pm \in \CPTP$, so one only needs to consider games such that $P(\Lambda,\G) \geq 0 \; \forall \Lambda \in \CPTP$.

On the other hand, by strong Lagrange duality (see App.~\ref{app:duals}) we can write
\begin{equation}\begin{aligned}\label{eq:rob_dual}
  R'(\Phi) + 1 = \max \lset \< W, J_\Phi \> \bar 0 \leq W \leq \rho \otimes \id_B,\; \rho \in \DD(A) \rset.
\end{aligned}\end{equation}
We can then make the following observations. Firstly, since the set of separable states in $\DD(A\otimes B)$ has a non-empty interior~\cite{zyczkowski_1998}, any Hermitian operator $X$ can be written as $X = \sum_{i=1}^{n} x_i\, \sigma_i \otimes \eta_i$ for some $\sigma_i \in \DD(A)$, $\eta_i \in \DD(B)$, $x_i \in \RR$, and $n \in \NN$. Then, choose the optimal $W$ in Eq.~\eqref{eq:rob_dual} and write $W^{T_A} = \sum_{i=1}^{n} x_i\, \sigma_i \otimes \eta_i$, where $T_A$ denotes the partial transpose. Defining the set $\{M_i\}_{i=1}^{n+1}$ by $M_i \coloneqq \eta_i / \norm{\sum_j \eta_j}{\infty}$ for $i\leq n$ and $M_{n+1} = \id - \sum_{i=1}^{n} M_i$, we have that
\begin{equation}\begin{aligned}
  W = \sum_i p_i w_i \, \sigma^T_i \otimes M_i
\end{aligned}\end{equation}
where $p_i = 1/n$ for $i \leq n$ and $p_{n+1} = 0$, and the coefficients $w_i$ are defined by $w_i = x_i n \norm{\sum_j \eta_j}{\infty}$. By the Choi-Jamiołkowski isomorphism and the linearity of $\Phi$, we then have for $\Phi$ that
\begin{equation}\begin{aligned}
  \< W, J_\Phi \> = \sum_i p_i w_i \< M_i, \Phi(\sigma_i) \> = P(\Phi, \G')
\end{aligned}\end{equation}
with $\G'$ defined by the above choices of $\{p_i, \sigma_i\}$, $\{M_i\}$, and $\{w_i\}$. Noticing that $W \geq 0 \Rightarrow P(\Gamma, \G') \geq 0 \; \forall \Gamma \in \CP$, this finally gives
\begin{equation}\begin{aligned}
\sup_{\G} \frac{P(\Phi, \G)}{\max_{\Lambda \in \CPTP} P(\Lambda, \G)}
 &\geq   \frac{P(\Phi, \G')}{\max_{\Lambda \in \CPTP} P(\Lambda, \G')}\\
 &\geq \frac{R'(\Phi) + 1}{{\max_{\Lambda \in \CPTP} R'(\Lambda) + 1}} \\
 &= R'(\Phi) + 1
\end{aligned}\end{equation}
where the second inequality follows since $P(\Phi, \G') = \<W, J_\Phi \> = R'(\Phi)+1$ holds by assumption while $P(\Lambda, \G') = \<W, J_\Lambda \> \leq R'(\Lambda)+1$ holds for any map $\Lambda$ by definition, and the last equality follows since $R'(\Lambda) = 0$ for any $\Lambda \in \CPTP$.
\end{proof}


\section{General bounds}\label{sec:bounds}

Useful bounds for the measures can be obtained by relating them with norms or quantities computed at the level of the Choi operator $J_\Phi$, avoiding an optimisation over all CPTNI or CPTP maps. For instance, the following relation with the trace norm generalises known bounds for the diamond norm~\cite{kliesch_2017,nechita_2018} (see also~\cite{jiang_2020}).
\begin{boxed}{white}
\begin{proposition}\label{prop:bounds_trace_norm}
For any Hermiticity-preserving map $\Phi \in \H(A,B)$, decompose $J_\Phi$ into its positive and negative parts as $J_\Phi = {J_\Phi}_+ - {J_\Phi}_-$ with ${J_\Phi}_\pm \geq 0$. Then
\begin{equation}\begin{aligned}
  \norm{J_\Phi}{1} &\geq \norm{\Phi}{\bnorm} \geq \frac{1}{d_A} \norm{J_\Phi}{1},\\
  \max \big\{ \Tr {J_\Phi}_+ - 1,\, \Tr {J_\Phi}_- \big\} &\geq R(\Phi) \geq \max \left\{ \frac{1}{d_A} \Tr {J_\Phi}_+ - 1,\,  \frac{1}{d_A} \Tr {J_\Phi}_- \right\},\\
  \Tr {J_\Phi}_+ - 1 &\geq R'(\Phi) \geq  \frac{1}{d_A} \Tr {J_\Phi}_+ - 1,\\
  \Tr {J_\Phi}_- &\geq R''(\Phi) \geq \frac{1}{d_A} \Tr {J_\Phi}_-.
\end{aligned}\end{equation}
\end{proposition}
\end{boxed}
\begin{proof}
Consider $\norm{\cdot}{\bnorm}$ first. Using the expression
\begin{equation}\begin{aligned}
  \norm{J_\Phi}{1} &= \min \lset \mu_+ + \mu_- \bar J_\Phi = \mu_+ \omega_+ - \mu_- \omega_-,\; \omega_{\pm} \in \DD(A \otimes B) \rset
\end{aligned}\end{equation}
we see that any such decomposition provides a feasible solution for $\norm{\Phi}{\bnorm}$, since $\omega_\pm$ constitute valid Choi operators of maps $\Omega_\pm \in \CPTN(A,B)$. The first inequality thus follows. The second inequality is a consequence of the bound $\norm{\Phi}{\bnorm} \geq \norm{\Phi}{\dia}$ from Cor.~\ref{cor:dia_cptni_ineq} and the fact that $\frac{1}{d_A} \norm{J_\Phi}{1}$ is known to lower bound the diamond norm (see e.g.~\cite{kliesch_2017,nechita_2018}). It can also be explicitly seen by noting that any decomposition of the form $J_\Phi = \lambda_+ J_{\Lambda_+} - \lambda_- J_{\Lambda_-}$ with $\Lambda_\pm \in \CPTNI$ can provide a decomposition for the trace norm by rescaling each $J_{\Lambda_\pm}$ by its trace; specifically,
\begin{equation}\begin{aligned}
\norm{J_{\Phi}}{1} \leq \lambda_+ \Tr J_{\Lambda_+} + \lambda_- \Tr J_{\Lambda_-},
\end{aligned}\end{equation}
and using the fact that $\Tr J_{\Lambda_\pm} \leq d_A \norm{J_{\Lambda_\pm}}{\infty} \leq d_A$ gives the desired bound.

The case of the robustness measures $R', R''$ follows analogously, where we now use the fact that $\Tr X _+ = \min \lset \mu \bar X \leq \mu \rho, \; \rho \in \DD \rset$ and $\Tr X_- = \min \lset \mu \bar X + \mu \rho \geq 0, \; \rho \in \DD \rset$ for any Hermitian $X$. For the robustness $R$, take $\lambda$ to be the greater of $\Tr J_{\Phi_+}-1$ and $\Tr J_{\Phi_-}$, and write
\begin{equation}\begin{aligned}
  J_{\Phi} + \lambda \frac{{J_\Phi}_-}{\lambda} = (1 + \lambda) \frac{{J_\Phi}_+}{1+\lambda}.
\end{aligned}\end{equation}
Since each $\frac{{J_\Phi}_\pm}{\lambda} \in \CPTNI$, this provides a valid feasible solution for $R$. On the other hand, $R(\Phi) \geq \max\{ R'(\Phi), R''(\Phi)\}$ by definition, from which the lower bound follows.
\end{proof}
Both the upper and the lower bounds in Prop.~\ref{prop:bounds_trace_norm} can be tight, as was shown already for the diamond norm~\cite{michel_2018}. However, better upper bounds can be obtained as follows.
\begin{boxed}{white}
\begin{proposition}\mbox{}\label{prop:bounds_upper}
For any Hermiticity-preserving map $\Phi \in \H(A,B)$, it holds that
\begin{equation}\begin{aligned}
  \norm{\Phi}{\dia} & \leq \lambda_{\max}(\Tr_B [{J_\Phi}_+ + {J_\Phi}_-]),\\
  \norm{\Phi}{\bnorm} &\leq \lambda_{\max}(\Tr_B {J_\Phi}_+) + \lambda_{\max}(\Tr_B {J_\Phi}_-),\\
    R(\Phi) &\leq \max \big\{ \lambda_{\max}(\Tr_B J_{\Phi_+}) - 1,\; \lambda_{\max}(\Tr_B J_{\Phi_-}) \big\},\\
  R'(\Phi) &\leq \lambda_{\max}(\Tr_B {J_\Phi}_+) -1,\\
  R''(\Phi) &\leq \lambda_{\max}(\Tr_B {J_\Phi}_-).\\
\end{aligned}\end{equation}
\end{proposition}
\end{boxed}
We note that the bound for the diamond norm, which we stated above for completeness, appeared previously in~\cite{nechita_2018}.
\begin{proof}
\br{The bounds for $\norm{\cdot}{\dia}$, $\norm{\cdot}{\bnorm}$, $R'$ and $R''$}  follow simply by using $J_\Phi = {J_\Phi}_+ - {J_\Phi}_-$ as feasible solutions \br{in the definitions.}

For the robustness $R$, take $\lambda$ to be the greater of $\lambda_{\max}(\Tr_B J_{\Phi_+})-1$ and $\lambda_{\max}(\Tr_B J_{\Phi_-})$, and write
\begin{equation}\begin{aligned}
  J_{\Phi} + \lambda \frac{{J_\Phi}_-}{\lambda} = (1 + \lambda) \frac{{J_\Phi}_+}{1+\lambda}.
\end{aligned}\end{equation}
\br{Since this is a feasible solution for $R$, we get $R(\Phi) \leq \lambda$.}
\end{proof}

As for lower bounds, we will first need to establish dual expressions for the considered measures. The following Proposition is an application of standard convex duality arguments, and we include details in Appendix~\ref{app:duals} for completeness.
\begin{boxed}{white}
\begin{proposition}\label{prop:duals}
For any $\Phi \in \H(A,B)$, the following dual expressions hold.
\begin{equation}\begin{aligned}\label{eq:duals}
  \norm{\Phi}{\dia} &= \max \lset  \< J_\Phi, W \> \bar - \rho \otimes \id_B \leq W \leq \rho \otimes \id_B,\; \rho \in \DD(A) \rset\\
  \norm{\Phi}{\bnorm} &= \max \lset  \< J_\Phi, W \> \bar - \rho \otimes \id_B \leq W \leq \sigma \otimes \id_B,\; \rho, \sigma \in \DD(A) \rset\\
  R(\Phi) &= \max \lset \< J_\Phi, W \> - \Tr Y \bar - X \otimes \id_B \leq W \leq Y \otimes \id_B,\; X, Y \geq 0,\; \Tr (X + Y) = 1 \rset\\
  R'(\Phi) &= \max \lset  \< J_\Phi, W \> - 1 \bar 0 \leq W \leq \rho \otimes \id_B,\; \rho \in \DD(A) \rset\\
  R''(\Phi) &= \max \lset  \< J_\Phi, W - \rho \otimes \id_B \> \bar 0 \leq W \leq \rho \otimes \id_B,\; \rho \in \DD(A) \rset\\
  &= \max \lset  \< J_\Phi, W \> - \Tr \Phi(\rho) \bar 0 \leq W \leq \rho \otimes \id_B,\; \rho \in \DD(A) \rset.\\
\end{aligned}\end{equation}
\end{proposition}
\end{boxed}

We can then obtain lower bounds by employing the dual optimisation problems. The bound for the diamond norm is well known~\cite{watrous_2004}, but we find it is insightful to rederive it using this approach\footnote{We remark the curious fact that, despite the apparent similarity, the bound for the diamond norm in Prop.~\ref{prop:bounds_lower} is not the induced Schatten norm $\norm{\cdot}{1\to 1}$, as the latter requires an optimisation over non-Hermitian input operators even when the map $\Phi$ is Hermiticity-preserving~\cite{watrous_2004}.}.
\begin{boxed}{white}
\begin{proposition}\label{prop:bounds_lower}
For any $\Phi \in \H(A,B)$ and any input state $\rho \in \DD(A)$, let $\Phi(\rho)_{\pm}$ denote the positive/negative part of the output operator $\Phi(\rho)$. Then
\begin{equation}\begin{aligned}
  \norm{\Phi}{\bnorm} \geq \norm{\Phi}{\dia} &\geq \max \lset \norm{\Phi(\rho)}{1} \bar \rho \in \DD(A) \rset\\
  &\geq \norm{\Tr_B J_\Phi}{\infty}\\
\norm{\Phi}{\bnorm} &\geq \max \lset \Tr \Phi(\rho)_- + \Tr \Phi(\sigma)_+ \bar \rho, \sigma \in \DD(A) \rset\\
&\geq \lambda_{\max} [(\Tr_B{J_\Phi})_+] + \lambda_{\max} [(\Tr_B{J_\Phi})_-]\\
  R(\Phi) &\geq \max \lset \Tr \Phi(X)_- + \Tr \Phi(Y)_+\textbf{} - \Tr Y \bar  X, Y \geq 0,\; \Tr (X + Y) = 1 \rset\\
  &\geq \max \Big\{ \lambda_{\max} [(\Tr_B{J_\Phi})_+] - 1,\, \lambda_{\max} [(\Tr_B{J_\Phi})_-] \Big\}\\
  R'(\Phi) &\geq \max \lset  \Tr \Phi(\rho)_+ - 1 \bar \rho \in \DD(A) \rset\\
  &\geq \lambda_{\max} [(\Tr_B{J_\Phi})_+] - 1\\
  R''(\Phi) &\geq \max \lset  \Tr \Phi(\rho)_- \bar  \rho \in \DD(A) \rset\\
  &\geq \lambda_{\max} [(\Tr_B{J_\Phi})_-].\\
\end{aligned}\end{equation}
\end{proposition}
\end{boxed}
\begin{proof}
Consider the diamond norm first. The main idea is to restrict the optimisation in the dual expression of $\norm{\cdot}{\dia}$ in \eqref{eq:duals} to operators of the form $W = \rho \otimes Z$ for some operator $Z$. Then we have
\begin{equation}\begin{aligned}
  \norm{\Phi}{\dia} &\geq \max \lset \< J_\Phi, \rho \otimes Z \> \bar - \id_B \leq Z \leq \id_B,\; \rho \in \DD(A) \rset\\
  &=  \max \lset \< \Phi(\rho^T), Z \> \bar - \id_B \leq Z \leq \id_B,\; \rho \in \DD(A) \rset\\
  &= \max \lset \norm{\Phi(\rho^T)}{1} \bar \; \rho \in \DD(A) \rset,
\end{aligned}\end{equation}
where the second line follows by the Choi-Jamiołkowski isomorphism. Taking $Z \in \{ \id, -\id \}$, we get the lower bound
\begin{equation}\begin{aligned}
  \norm{\Phi}{\dia} &\geq \max \lset \pm \< \Tr_B J_\Phi, \rho \> \bar \rho \in \DD(A) \rset\\
  &= \norm{\Tr_B J_\Phi}{\infty}.
\end{aligned}\end{equation}

In the case of $\norm{\cdot}{\bnorm}$, we use feasible solutions of the form $W = \rho \otimes Z + \sigma \otimes V$ with $-\id \leq Z \leq 0$ and $0 \leq V \leq \id$ to obtain the stated bound analogously --- the crucial observation being that $\max \lset \< A, B \> \bar 0 \leq B \leq \id \rset = \Tr A_+$ for any Hermitian $A$. The other measures follow in the same way.
\end{proof}
Note the similarity between the eigenvalue-based lower bounds of Prop.~\ref{prop:bounds_lower} and the upper bounds of Prop.~\ref{prop:bounds_upper}: the upper bounds consider the eigenvalues after decomposing $J_\Phi$ as ${J_\Phi}_+ - {J_\Phi}_-$, while the lower bounds use the positive and negative parts of $\Tr_B J_\Phi$.

An immediate consequence is that for any completely positive map $\Phi$, it holds that
\begin{equation}\begin{aligned}\label{eq:equality_CP}
  \norm{\Phi}{\bnorm} = \norm{\Phi}{\dia} = \norm{\Tr_B J_\Phi}{\infty}
\end{aligned}\end{equation}
since the operators $J_\Phi$ and  $\Tr_B J_\Phi$ are both positive semidefinite. However, the lower bounds allow us to show explicitly that the equality $\norm{\Phi}{\bnorm} = \norm{\Phi}{\dia}$ is no longer true for maps which are neither CP nor trace preserving, and in fact the extreme disparity of $\norm{\Phi}{\bnorm} = 2 \norm{\Phi}{\dia}$ (cf.\ Cor.~\ref{cor:dia_cptni_ineq}) can be achieved. Consider for instance the case when 
\begin{equation}\begin{aligned}
\Phi(\cdot) = \braket{0|\cdot|0} \proj{0} - \braket{1|\cdot|1} \proj{1}.
\end{aligned}\end{equation}
Decomposing $J_{\Phi} = \proj{0} \otimes \proj{0} - \proj{1} \otimes \proj{1}$ into its positive and negative parts, the bound of Prop.~\ref{prop:bounds_upper} gives $\norm{\Phi}{\dia} \leq 1$. However, the best upper bound we get for $\norm{\Phi}{\bnorm}$ is 2, and it is indeed tight: we have $\Phi(\proj{0}) = \proj{0}$ and $\Phi(\proj{1}) = - \proj{1}$, and so Prop.~\ref{prop:bounds_lower} gives $\norm{\Phi}{\bnorm}\geq 2$. A similar argument can be used to show that $R(\Phi) = 1$, which in particular implies that $2 R(\Phi) + 1 > \norm{\Phi}{\bnorm} > \norm{\Phi}{\dia}$.

All of the bounds that we established in this section can be tight, as we shall demonstrate in what follows.


\section{Applications and examples}\label{sec:applications}

\subsection{Positive maps and structural physical approximation}

Positive maps constitute a fundamental way to detect and characterise quantum entanglement~\cite{horodecki_1996-1,guhne_2009,horodecki_2009}. One of the most studied approaches to implementing such maps in practice is the structural physical approximation (SPA)~\cite{horodecki_2003-4,horodecki_2002-1}, which aims to approximate a given positive map $\Phi$ with a physical quantum channel by considering decompositions of the form $\Phi + \varsigma \D$, where $\D$ is the completely depolarising channel, $J_\D = \id / d_B$. Such approximations have found use in both understanding the properties of positive maps~\cite{korbicz_2008,shultz_2015}, as well as in realising them in experiments~\cite{horodecki_2002-1,lim_2011,bae_2017}.

Intuitively, the robustness measures can then be understood as different approaches to defining an optimised SPA to the map $\Phi$, by allowing channels other than the depolarising map to be used in the decomposition (cf.~\cite{jiang_2020}). We will now discuss the similarities and differences between the approaches by studying two representative examples of positive maps.

\paragraph{Transposition map.} Consider first the transposition map $T \in \H(A,A)$. Letting $\SPA(T)$ denote the minimal amount $\varsigma$ needed for $(T + \varsigma \D)/(1+\varsigma)$ to be a quantum channel, it can be easily verified that $\SPA(T) = d_A$. However, by making a more suitable choice of a channel in the optimisation, our robustness measures construct an approximation as $(T + \lambda \Lambda)/(1+\lambda)$ where $\lambda = \frac{1}{2}(d_A-1)$ already suffices to ensure that this is a valid physical channel. From this we see that $R(T) =\frac{1}{2}(d_A-1)$ and hence $\norm{T}{\bnorm} = d_A$. Quantitatively, the advantage gained by allowing arbitrary channels in such decompositions can therefore be significant.

To understand why a better approximation can be obtained, let us take a closer look at the optimal decomposition for this map. Our generalised approach can take into consideration the fact that the Choi operator of the transposition map, $J_T$ (the swap operator), already has a non-trivial positive part, which means that there is no need to act on that part of the space. More specifically, a better approximation is obtained simply by defining the map $\displaystyle J_\Lambda = \frac{\id - J_T}{d_A - 1} \in \CPTP$ and mixing as 
\begin{equation}\begin{aligned}
  J_T + \frac12 (d_A - 1) J_\Lambda \geq 0 \;\Rightarrow\; \frac{2}{d_A+1}\, T + \frac{d_A-1}{d_A+1} \,\Lambda \in \CPTP.
\end{aligned}\end{equation}
Structurally, this is not too different from the SPA --- the only maps involved in the combination are the depolarising channel and the transposition map itself, even if the optimal approximation is not simply a convex mixture of the two. Indeed, we could define an optimised structural physical approximation which allows for such decompositions to be used:
\begin{equation}\begin{aligned}\label{eq:spa_modified}
  \SPA'(\Phi) \coloneqq& \min \lsetr \varsigma' \barr J_{\Phi} + \varsigma' \left[ \frac{\lambda_{\max}(J_\Phi) \id - J_\Phi}{\lambda_{\max}(J_\Phi) d_B - 1}\right] \geq 0 \rsetr\\
=& - \lambda_{\min}(J_\Phi) \frac{d_B \lambda_{\max}(J_\Phi) - 1}{\lambda_{\max}(J_\Phi) - \lambda_{\min}(J_\Phi)},
\end{aligned}\end{equation}
with the expression valid for any map such that $\lambda_{\min}(J_{\Phi}) < \lambda_{\max}(J_\Phi) \neq d_B^{-1}$. 
This can be used to give a general bound to the robustness measures.
\begin{boxed}{white}
\begin{proposition}
For any trace-preserving map $\Phi \in \H(A,B)$ such that $\Phi \neq \D$, it holds that
\begin{equation}\begin{aligned}
    R(\Phi) \leq \SPA'(\Phi) \leq \SPA(\Phi).
\end{aligned}\end{equation}
\end{proposition}
\end{boxed}
In the case of the transpose, it holds that $\SPA'(T)=R(T)=\frac{1}{2}(d_A-1)$, so we know that an optimal approximation of the transposition map can be realised with only the depolarising channel, as long as one considers the optimised approach of Eq.~\eqref{eq:spa_modified}. However, this is not the case for general maps, and the advantages offered by the generalised robustness approach can provide new insight into optimal approximations of maps, as we shall see in the following.

\paragraph{Choi map.} The Choi map $\C \in \H(A,A)$ with $d_A = 3$ is an example of an indecomposable positive map, and is defined by~\cite{choi_1980}
\begin{equation}\begin{aligned}
  \C(X) \coloneqq \begin{pmatrix} X_{11}+X_{22} & -X_{12} & -X_{13} \\ -X_{21} & X_{22}+X_{33} & -X_{23} \\ -X_{31} & -X_{32} & X_{33} + X_{11} \end{pmatrix}
\end{aligned}\end{equation}
where $X_{ij}$ denote the matrix elements of $X$ in a chosen basis. \br{A numerical evaluation shows} that the optimal decompositions for $\C$ give $\SPA(\C) = \frac{3}{2}$ and $\SPA'(\C) = \frac{2}{3}$. With the robustness, an improved choice can be obtained by choosing $\Lambda = \idc$ and mixing as $J_\C + \frac{1}{6} J_{\idc} \geq 0$, yielding $R(\C) = \frac{1}{6}$. Consequently, mixing with more general maps can not only provide quantitative improvements, but also identify ways of implementing non-CPTP maps which are impossible to find with the standard structural physical approximations.

An interesting difference between the SPA- and robustness-based approaches is that the optimal SPA of the Choi map is a measure-and-prepare (entanglement-breaking) channel~\cite{korbicz_2008}, while the map obtained in the robustness-based approach is not (as can be verified with the PPT criterion). Since measure-and-prepare channels enjoy an easy implementation in practical settings, it would be an interesting extension of our approach to consider the extent of a quantitative advantage that can be maintained while requiring that the optimal CPTP approximation be entanglement breaking.

We also note that another approach to realising positive maps was studied in Ref.~\cite{dong_2019} by using multiple copies of the input state, where a related SPA-based approximation was also considered. An extension of the methods of our work to this framework could provide additional insight into the implementability of positive maps.


\subsection{Inverse quantum channels}\label{sec:examples_inverses}

A fundamentally important case of a non-CPTP map encountered in many settings is the inverse linear map of a bijective quantum channel, that is, a map such that $\Lambda^{-1} \circ \Lambda = \Lambda \circ \Lambda^{-1} = \idc$.\footnote{We note that in many cases it suffices to consider only left or right inverses, but we assume two-sided invertibility for simplicity.} Note that such an inverse is not guaranteed to exist for a general channel, and even when it does, it will not form a valid quantum channel unless $\Lambda$ is a unitary map. However, many important cases of quantum dynamics are indeed invertible, allowing us to study their inverses in the formalism of our work.

\paragraph{Non-Markovianity.} One setting in which channel inverses play a role is the study of non-Markovianity. Among the different ways to define Markovian evolution, a common way is to say that a time-dependent evolution governed by the channel $\Lambda_{t,0}$ is Markovian if it behaves as a physical map over any time interval $[t,t+\delta t]$. Mathematically, any $\Lambda_{t,0}$ satisfying this condition is said to be CP-divisible~\cite{rivas_2010,chruscinski_2014,rivas_2014}, which can be formalised by the statement that for all times $t$ and $s \leq t$ we can write
$$\Lambda_{t,0} = \Xi_{t,s} \circ \Lambda_{s,0}$$
where the propagator $\Xi_{t,s}$ is a CPTP map. For more general channels, the decomposition $\Lambda_{t,0} = \Xi_{t,s} \circ \Lambda_{s,0}$ results in some $\Xi_{t,s}$ that is non-CPTP, 
indicating that Markovian dynamics break down after some time point $s$.

Observe that, provided $\Lambda_{t,0}$ is invertible for all $t$, we can take $\Xi_{t,s} = \Lambda_{t,0} \circ \Lambda^{-1}_{s,0}$. Therefore, \br{the non-physicality of $\Lambda_{t,0} \circ \Lambda^{-1}_{s,0}$ serves as an indicator of non-Markovianity, and --- since this map is trace preserving for any trace-preserving $\Lambda$ --- the diamond norm $\norm{\Lambda_{t,0} \circ \Lambda^{-1}_{s,0}}{\dia}$ can be used as a quantitative measure of  non-Markovianity over the time-interval $[t,s]$}. This is similar to the original approach of Ref.~\cite{rivas_2010} where a quantifier based on the trace norm of the Choi operator was employed --- the advantage of our definition is the ability to interpret this quantity operationally.

Specifically, we observe that quantum mechanics is ultimately a Markovian theory: if we had knowledge of all relevant objects, then all quantum dynamics could be described by Markovian unitary dynamics. That is, any information from the past that is relevant to the future must pass through the present, and hence the optimal prediction of future observational statistics ultimately depends only on the the present state of reality. Non-Markovianity is an artefact of not tracking all relevant information in the present. In our context, this arises as our mathematical characterisation of the candidate channel, $\Lambda_{s,0}$, does not track the state of the environment. 
The operational relevance of $\norm{\Lambda_{t,0} \circ \Lambda^{-1}_{s,0}}{\dia}$ then becomes more evident. Notably, in Sec.~\ref{sec:simulation} we presented a systematic means of simulating any unphysical map $\Xi_{t,s}$ by introducing an ancillary system $X$. Here, we may think of this as building a Markovian model for $\Xi_{t,s}$ by introducing $X = \sum_i \mu_i \rho_i$ as an ``artificial environment''. The feeding in of different states $\rho_i$ depending on $X$ then represents a means in which non-Markovian behaviour on the system is realised. While this construction does not immediately look physical (as it allows affine mixtures of quantum states), it can be simulated by a classical computer with sufficient resource overhead. The resource costs of doing so --- $\norm{\Xi_{t,s}}{\dia}$ --- thus represents a bound on the information processing capabilities of the environment that enable said non-Markovian behaviour to emerge. 

There are multiple approaches for extending this to a time-independent measure of non-Markovianity of $\Lambda$. One could, for example, take the supremum of the measure $\norm{\Lambda_{t,0} \circ \Lambda^{-1}_{s,0}}{\dia}$ over all $t$ and $s$. This would then characterise how much extra information processing we need beyond tracking the state of the system at time $s$ to simulate dynamics over the time-interval
$[s,t]$. 
We may also follow an approach based on Ref.~\cite{rivas_2010} and define $\displaystyle \mathcal{I}_{\,\dia} (\Lambda) \coloneqq \int_{0}^\infty g_{\,\dia,t}(\Lambda) \,\mathrm{d}t$, where $g_{\,\dia,t}$ can be understood as the  right-hand derivative of the diamond norm of the dynamics at time $t$:
\begin{equation}\begin{aligned}
  g_{\,\dia,t}(\Lambda) \coloneqq \lim_{\ve \to 0^+} \frac{\norm{\Lambda_{t+\ve,0} \circ \Lambda_{t,0}^{-1}}{\dia}-\norm{\Lambda_{t,0} \circ \Lambda_{t,0}^{-1}}{\dia}}{\ve} = \lim_{\ve \to 0^+} \frac{\norm{\Lambda_{t+\ve,0} \circ \Lambda_{t,0}^{-1}}{\dia}-1}{\ve}.
\end{aligned}\end{equation}
$\mathcal{I}_{\,\dia} (\Lambda)$ therefore represents the total amount of non-Markovianity in this evolution. A suitable normalisation of this quantity can allow for the comparison of the strength of non-Markovianity in different settings~\cite{rivas_2010,rivas_2014}. We leave a careful consideration of these possibilities to future work.

\paragraph{Error mitigation.} Another application for the study of channel inverses is error mitigation. 
This setting considers the scenario where one is tasked with computing expectation values of the type $\Tr[ \U(\rho) A ]$ for an input state $\rho$, ideal gate $\U$, and observable $A$, while operations are followed by a noise channel $\Theta$.
A leading approach to this problem, called probabilistic error cancellation~\cite{temme_2017,endo_2018}, is to counteract the noise with the inverse map $\Theta^{-1}$, so that 
$\Tr[\U(\rho) A]=\Tr[ \Theta \circ \Theta^{-1}\circ\U(\rho) A ]$. 
By decomposing $\Theta^{-1}$ into a quasiprobability distribution over a convex subset of channels $\P=\{\Lambda_i\}$ such that $\Lambda_i\circ\U$ would be implementable on a (fictitious) noiseless device, standard quasiprobability sampling arguments allows us to construct an unbiased estimator for $\Tr[\U(\rho) A]$ using only operations implementable on a noisy device.
The optimal overhead cost of such a procedure scales as $\gamma_\P(\Theta)^2$, where~\cite{temme_2017,takagi_2020-2}
\begin{equation}\begin{aligned}
  \gamma_\P(\Theta) &= \min \lset \sum_i |\lambda_i| \bar \Theta^{-1} = \sum_i \lambda_i \Lambda_i,\; \Lambda_i \in \P \rset\\
  &= \min \lset \lambda_+ + \lambda_- \bar \Theta^{-1} = \lambda_+ \Lambda_+ - \lambda_- \Lambda_-,\; \Lambda_\pm \in \P \rset.
  \label{eq:sampling cost}
\end{aligned}\end{equation}
The specific choice of $\P$ can be made depending on not only the physical setting in consideration, but also on one's precise motivations. On the one hand, a set with a finite number of operations (e.g., Clifford gates) turns Eq.~\eqref{eq:sampling cost} into a linear program~\cite{temme_2017,endo_2018}, making the overhead cost easily computable while sacrificing the expressibility of devices.
On the other hand, choosing a larger set with an infinite number of implementable operations takes into account a larger expressibility~\cite{takagi_2020-2}, but makes the computation of Eq.~\eqref{eq:sampling cost} hard in general.   
Here, to accommodate  computability and expressibility at the same time, we take another approach considered in Ref.~\cite{jiang_2020,Xiong2020sampling}: we choose $\P$ to be \emph{all} physical quantum channels. 
We notice that the norm $\norm{\cdot}{\bnorm}$ provides the cost of error mitigation in this setting as $\gamma_{\CPTP} (\Theta) = \norm{\Theta^{-1}}{\bnorm} = \norm{\Theta^{-1}}{\dia}$, which can be efficiently computed by semidefinite programming. 
Although this choice of $\P$ might seem too permissive, the lower bound obtained through this approach can actually match known achievability results (upper bounds)~\cite{jiang_2020}, showing new optimality results and even improving on the specialised characterisation of Ref.~\cite{takagi_2020-2} in some cases. Of note is the fact that, since any inverse map $\Theta^{-1}$ of a quantum channel $\Theta$ is trace preserving, our Thm.~\ref{thm:rob_diamond} shows a new application of the diamond norm in bounding the cost of error mitigation: it always holds that $\gamma_\P(\Theta) \geq \norm{\Theta^{-1}}{\dia}$, regardless of the choice of $\P$.

In some cases --- such as when experiencing the leakage or loss of some qubits during computation --- the noisy evolution can actually correspond to a map which is not trace preserving. Although many previous approaches did not take this into consideration, our methods explicitly extend to such maps, allowing one to understand the simulation of non-trace-preserving linear maps through Thm.~\ref{thm:cost}. Related settings which our methods can characterise include the so-called linear quantum error correction~\cite{shabani_2009}, which aims to correct errors of systems undergoing general, non-CPTP dynamics $\Theta$, as well as error mitigation for non-Markovian noise~\cite{Hakoshima2021nonMarkovian}, where the mitigation cost can be related to a measure of non-Markovianity. In such cases, our approach can thus help understand the implementation of not only the inverse maps, but also the dynamics themselves.


\subsubsection{Computing the measures}

To showcase the application of our methods and evaluate the measures for some representative examples, we will consider the inverse maps of several fundamental types of noisy quantum evolutions: depolarising, amplitude damping, dephasing, and qubit leakage channels. The expressions for the first two appeared in Ref.~\cite{jiang_2020}, which we rederive using the methods and results of this work. 
We also find for the first three that the optimal decomposition into $\Lambda_\pm$ for the norm $\|\Theta^{-1}\|_{\bnorm}$ (Eq.~\eqref{eq:CPTNI norm def}), can be taken as convex mixtures of unitaries and state preparations. Thus, $\|\Theta^{-1}\|_{\bnorm}$ also serves as the optimal cost $\gamma_\P(\Theta^{-1})$ with a smaller set $\P$ as considered in Ref.~\cite{takagi_2020-2}, indicating that the capability to implement all CPTNI maps does not provide any advantage over that of implementing unitaries and state preparations only.  
Note that the inverses of trace-preserving maps are trace preserving, and so in such cases the equality $\norm{\Phi}{\bnorm} = \norm{\Phi}{\dia} = 2R(\Phi) + 1$ holds by Thm.~\ref{thm:rob_diamond}, which means that it will suffice to evaluate any one of the measures.

\paragraph{Depolarising noise.} The depolarising channel, given by $\D_p(X) \coloneqq (1-p) X + p \Tr X \frac{\id}{d_A}$ for some noise parameter $p\in[0,1)$, has the inverse $\D^{-1}_p(X) = \frac{1}{1-p} X - \frac{p}{1-p} \Tr X \frac{\id}{d_A}$. This gives
\begin{equation}\begin{aligned}
  J_{\D^{-1}_p} = \frac{1}{1-p} \proj{\Omega} - \frac{p}{(1-p)d_A} \id_{A \otimes A}.
\end{aligned}\end{equation}
Importantly, one can notice that $\Tr_B {J_{\D^{-1}_p}}_+$ and $\Tr_B {J_{\D^{-1}_p}}_-$ are proportional to identity. As first noticed in~\cite{nechita_2018,michel_2018}, this means that the lower bound $\frac{1}{d_A} \norm{J_{\D^{-1}_p}}{1}$ of Prop.~\ref{prop:bounds_trace_norm} matches the upper bound $\lambda_{\max} \left(\Tr_B \left[ {J_{\D^{-1}_p}}_+ + {J_{\D^{-1}_p}}_- \right]\right)$ of Prop.~\ref{prop:bounds_upper}\footnote{In fact, $\norm{\Phi}{\dia} = \frac{1}{d_A} \norm{J_\Phi}{1}$ if and only if $\Tr_B \left({J_{\Phi}}_+ + {J_{\Phi}}_-\right) \propto \id$~\cite{nechita_2018,michel_2018}.}.
We thus get
\begin{equation}\begin{aligned}
  \norm{\D^{-1}_p}{\bnorm} = \norm{\D^{-1}_p}{\dia} &= \frac{1}{d_A} \norm{J_{\D^{-1}_p}}{1} = \frac{1 + \left(1-2d_A^{-2}\right) p}{1-p}.
\end{aligned}\end{equation}

\paragraph{Dephasing noise.} The generalised dephasing channel~\cite{devetak_2005-2} is defined by $\Delta_{\textbf{p}}(X) \coloneqq \sum_{i=0}^{d_A-1} p_i Z_i X Z_i^\dagger$, where $\textbf{p} = (p_0, \ldots, p_{d_A-1})$ is a chosen set of noise parameters $p_i \geq 0$, and $Z_i$ refers to the qudit clock operators
\begin{equation}\begin{aligned}
  Z_i = \sum_{j=0}^{d_A-1} \omega^{ij} \proj{j}
\end{aligned}\end{equation}
in some basis  $\{\ket{i}\}$, with $\omega$ being a primitive $d_A$th root of unity. In the case of $d_A = 2$, this recovers the usual qubit dephasing channel $\Delta_{p}(X) = (1-p) X + p Z X Z^\dagger$. One can notice that the action of this channel can be represented by $\Delta_{\textbf{p}} (X) = X \odot S$ where $\odot$ denotes the element-wise matrix product (Schur/Hadamard product), and
\begin{equation}\begin{aligned}
  (S)_{jk} = \sum_{i=0}^{d_A-1} p_i \omega^{ij} (\omega^{ik})^* = \sum_{i=0}^{d_A-1} p_i \omega^{i(j-k)} \qquad j,k = 0, \ldots d_A-1
\end{aligned}\end{equation}
in the same basis $\{\ket{i}\}$.
Provided that the coefficients of $S$ are non-zero (that is, $\Delta_{\textbf{p}}$ does not act as a completely dephasing channel on any subspace), the map is invertible as $\Delta^{-1}_{\textbf{p}}(X) = X \odot \overline{S}$ with $\overline{S}$ defined by
\begin{equation}\begin{aligned}
  (\overline{S})_{jk} = \frac{1}{(S)_{jk}} \qquad j,k = 0, \ldots d_A-1.
\end{aligned}\end{equation}
We will now show that $\norm{\Delta^{-1}_{\textbf{p}}}{\bnorm} = \norm{\Delta^{-1}_{\textbf{p}}}{\dia} = \frac{1}{d_A} \norm{J_{\Delta^{-1}_{\textbf{p}}}}{1} = \frac{1}{d_A}\norm{\overline{S}}{1}$.

The equality $\norm{\Delta^{-1}_{\textbf{p}}}{\bnorm} = \norm{\Delta^{-1}_{\textbf{p}}}{\dia}$ is a consequence of Thm.~\ref{thm:rob_diamond}; note here that we do not actually need to impose that $\Delta_{\textbf{p}}$ be trace preserving (i.e., that $\sum_i p_i = 1$), since both $\Delta_{\textbf{p}}$ and $\Delta^{-1}_{\textbf{p}}$ are always proportional to a trace-preserving map by construction.

To show the equality $\norm{\Delta^{-1}_{\textbf{p}}}{\bnorm} = \frac{1}{d_A}\norm{\overline{S}}{1}$, consider the decomposition of $\overline{S}$ as $\overline{S} = \overline{S}_+ - \overline{S}_-$. Crucially, since $S$ is a circulant matrix, so is $\overline{S}$, and hence it can be diagonalised by the Fourier transform matrix $(F)_{jk} = \frac{1}{\sqrt{d_A}} \omega^{jk}$~\cite[2.2.P10]{horn_2012}. Each eigenvector of $\overline{S}$ is therefore of the form
\begin{equation}\begin{aligned}
  \ket{s_m} = \frac{1}{\sqrt{d_A}} \sum_{i=0}^{d_A - 1} \omega^{im} \ket{i},
  \label{eq:dephasing eigenstate}
\end{aligned}\end{equation}
ensuring in particular that all diagonal elements of each density matrix $\proj{s_m}$ are equal. This entails that $\overline{S}_+$ and $\overline{S}_-$ both have constant diagonals. Define now the maps
\begin{equation}\begin{aligned}
  \Lambda_\pm (X) \coloneqq X \odot \overline{S}_\pm.
\end{aligned}\end{equation}
Since $\overline{S}_\pm \geq 0$, each such map is completely positive~\cite[Thm. 3.7]{paulsen_2002}, and clearly $\Lambda'_\pm \coloneqq \Lambda_\pm d_A / \Tr(\overline{S}_\pm)$ is trace preserving as we have just seen that $(\overline{S}_\pm)_{ii} = (\overline{S}_\pm)_{jj} \; \forall i,j$. Thus we have a decomposition as
\begin{equation}\begin{aligned}
  \Delta^{-1}_{\textbf{p}} = \frac{\Tr \overline{S}_+}{d_A} \Lambda'_+ - \frac{\Tr \overline{S}_-}{d_A} \Lambda'_-, \quad \Lambda'_\pm \in \CPTP,
 \label{eq:dephasing inverse decomposition}
\end{aligned}\end{equation}
from which we get the bound $\norm{\Delta^{-1}_{\textbf{p}}}{\bnorm} \leq \frac{1}{d_A}\norm{\overline{S}}{1}$. On the other hand, let $\ket{\psi} = \frac{1}{\sqrt{d_A}} \sum_{i=0}^{d_A-1} \ket{i}$ and use Prop.~\ref{prop:bounds_lower} to get
\begin{equation}\begin{aligned}
  \norm{\Delta^{-1}_{\textbf{p}}}{\bnorm} \geq \norm{\Delta^{-1}_{\textbf{p}}(\proj{\psi})}{1} = \frac{1}{d_A}\norm{\overline{S}}{1}.
\end{aligned}\end{equation}
Finally, the equality $\norm{J_{\Delta^{-1}_{\textbf{p}}}}{1} = \norm{\overline{S}}{1}$ is obtained by noticing that $J_{\Delta^{-1}_{\textbf{p}}} = \sum_{i,j} (\overline{S})_{ij} \ketbra{ii}{jj}$ which has the same eigenvalues as $\overline{S}$.

The eigenvalues of $\overline{S}$ can be readily obtained due to the fact that it is a circulant matrix~\cite[2.2.P10]{horn_2012}, allowing for a straightforward computation of the trace norm $\norm{\overline{S}}{1}$ and altogether giving
\begin{equation}\begin{aligned}
  \norm{\Delta^{-1}_{\textbf{p}}}{\dia} = \frac{1}{d_A} \sum_{m=0}^{d_A-1} \left| \sum_{j=0}^{d_A-1} \left( \sum_{i=0}^{d_A-1} p_i \omega^{j(i-m)} \right)^{-1} \right|.
\end{aligned}\end{equation}

For the qubit dephasing channel with $p\in[0,\frac{1}{2})$, we recover
\begin{equation}\begin{aligned}
  \norm{\Delta^{-1}_p}{\dia} = \frac{1}{2} \norm{\begin{pmatrix}1 & \frac{1}{1-2p} \\\frac{1}{1-2p} & 1\end{pmatrix}}{1} = \frac{1}{1-2p}.
\end{aligned}\end{equation}

Since each eigenvector $\ket{s_m}$ for $\overline{S}$ in \eqref{eq:dephasing eigenstate} corresponds to the application of $Z_m$, $\Lambda'_\pm$ in \eqref{eq:dephasing inverse decomposition} are realised as probabilistic applications of the generalised phase unitaries.

\paragraph{Amplitude damping noise.} The qubit amplitude damping channel $\A_\gamma(\cdot) = A_0 \cdot A_0^\dagger + A_1 \cdot A_1^\dagger$ is defined by the Kraus operators $A_0 \coloneqq \proj{0} + \sqrt{1-\gamma} \proj{1}$ and $A_1 \coloneqq \sqrt{\gamma} \ketbra{0}{1}$. Using the fact that
\begin{equation}\begin{aligned}
  \proj{1} &= \frac{1}{1-\gamma} \A_\gamma(\proj{1}) - \frac{\gamma}{1-\gamma} \proj{0}\\
  &= \frac{1}{1-\gamma} \A_\gamma(\proj{1}) - \frac{\gamma}{1-\gamma} \A_\gamma(\proj{0}),
  \end{aligned}\end{equation}
we have
\begin{equation}\begin{aligned}
  \A_\gamma^{-1}(\proj{1}) = \frac{1}{1-\gamma} \proj{1} - \frac{\gamma}{1-\gamma} \proj{0}.
\end{aligned}\end{equation}
Proposition \ref{prop:bounds_lower} thus gives
\begin{equation}\begin{aligned}
  \norm{\A^{-1}_\gamma}{\bnorm} = \norm{\A^{-1}_\gamma}{\dia} &\geq \norm{\A^{-1}_\gamma(\proj{1})}{1}\\
  &= \frac{1+\gamma}{1-\gamma}.
\end{aligned}\end{equation}
A matching upper bound can be obtained by explicitly computing $J_{\A_\gamma^{-1}}$ (see e.g.~\cite{temme_2017,takagi_2020-2}) and using the upper bound in Prop.~\ref{prop:bounds_upper}.

The above shows a rather general method of obtaining lower bounds for linear maps which are inverses of other linear maps, without having to explicitly compute the full inverse map. Indeed, this can be extended to maps which only approximately invert a given channel --- useful, for instance, when dealing with non-invertible maps, or when aiming to reduce the cost of implementing a given map by only requiring that it approximately mitigates the error.

\begin{boxed}{white}
\begin{proposition}
Let $\Phi \in \H(A,B)$ and $\wt\Phi \in \H(B,A)$ be such that $\snorm{ \wt\Phi \circ \Phi(\rho) - \rho }{1} \leq \ve$ for all $\rho \in \DD(A)$. Then
\begin{equation}\begin{aligned}
  \snorm{\wt\Phi}{\dia} &\geq \max \lset \norm{Z}{1}(1-\ve) \bar \Phi(Z) \in \DD(B) \rset \\
\snorm{\wt\Phi}{\bnorm} &\geq \max \lset \Tr Z_- + \Tr Q_+ - \ve (\norm{Z}{1}+\norm{Q}{1}) \bar \Phi(Z), \Phi(Q) \in \DD(B) \rset \\
  R(\wt\Phi) &\geq \max \lset \Tr Z_- + \Tr Q_+ - \Tr \Phi(Q) - \ve (\norm{Z}{1}+\norm{Q}{1}) \bar \Phi(Z), \Phi(Q) \geq 0, \right.\\
  &\hphantom{\geq \max \lset \Tr Z_- + \Tr Q_+ - \Tr \Phi(Q) - \ve (\norm{Z}{1}+\norm{Q}{1}) \bar \right.} \Tr \Phi(Z + Q) = 1 \big\}\\
  R'(\wt\Phi) &\geq \max \lset  \Tr Z_+ - 1 - \ve \norm{Z}{1} \bar \Phi(Z) \in \DD(B) \rset\\
  R''(\wt\Phi) &\geq \max \lset  \Tr Z_- - \ve \norm{Z}{1} \bar  \Phi(Z) \in \DD(B) \rset.\\
\end{aligned}\end{equation}
\end{proposition}
\end{boxed}
\begin{proof}
We use Prop.~\ref{prop:bounds_lower} to get that
\begin{equation}\begin{aligned}
  \snorm{\wt\Phi}{\dia} &\geq \max \lset \snorm{\wt\Phi(\sigma)}{1} \bar \sigma \in \DD(B) \cap \mathrm{ran}(\Phi) \rset\\
  &= \max \lset \snorm{\wt\Phi \circ \Phi (Z)}{1} \bar \Phi(Z) \in \DD(B) \rset\\
  &\geq \max \lset \norm{Z}{1} - \snorm{ Z - \wt\Phi \circ \Phi (Z) }{1} \bar \Phi(Z) \in \DD(B) \rset\\
  &\geq \max \lset \norm{Z}{1} (1-\ve) \bar \Phi(Z) \in \DD(B) \rset.
\end{aligned}\end{equation}
The third line follows by the triangle inequality, and the last line is a consequence of the assumption that $\snorm{ \wt\Phi \circ \Phi(\rho) - \rho }{1} \leq \ve$ for all $\rho \in \DD(A)$, since we can write any $Z = \mu_+ \rho_+ - \mu_- \rho_-$ for some $\rho_\pm \in \DD(A)$ to get $\norm{\wt\Phi \circ \Phi(Z) - Z}{1} \leq \ve (\mu_+ + \mu_-) \leq \ve \norm{Z}{1}$. The case of the other measures is analogous: using the variational form of the function $\Tr Z_+$ (and similarly $\Tr Z_-$) we can obtain
\begin{equation}\begin{aligned}
  \Tr \wt\Phi(\sigma)_+ &= \max \lset \< \wt\Phi \circ \Phi (Z), W \> \bar 0 \leq W \leq \id \rset\\
  &= \max \lset \< Z , W \> - \< Z - \wt\Phi \circ \Phi(Z), W \> \bar 0 \leq W \leq \id \rset\\
  &\geq \max \lset \< Z , W \> - \snorm{Z - \wt\Phi\circ \Phi(Z)}{1} \norm{W}{\infty} \bar 0 \leq W \leq \id \rset\\
  &\geq \Tr Z_+ - \ve \norm{Z}{1}
\end{aligned}\end{equation}
where we used the Cauchy-Schwarz inequality. Using these bounds in Prop.~\ref{prop:bounds_lower} yields the stated result.
\end{proof}

\paragraph{Leakage error.} Consider the qubit leakage error $\L_p(\cdot)=L_p\cdot L_p^\dagger$ where $L_p\coloneqq \dm{0}+\sqrt{1-p}\dm{1}$. 
This represents a situation where the excited state is lost with probability $1-p$, and this stochastic nature is reflected to the fact that $\L_p$ is not trace preserving.
The inverse of the leakage error is given by $\L_p^{-1}(\cdot) = L_p^{-1} \cdot L_p^{-1}$. Since this is a completely positive map, Eq.~\eqref{eq:equality_CP} gives
\begin{equation}\begin{aligned}
  \norm{\L_p^{-1}}{\bnorm} = \norm{\L_p^{-1}}{\dia} = \norm{\Tr_B J_{\L_p^{-1}}}{\infty} = \frac{1}{1-p}.
\end{aligned}\end{equation}
Note that the inverse can be realised as 
\bal
\L_p^{-1}=\frac{1}{2}\left(1+\frac{1}{\sqrt{1-p}}\right)\idc-\frac{1}{2}\left(\frac{1}{\sqrt{1-p}}-1\right)\Z+\frac{p}{1-p}\Pi_{\dm{1}}
\eal
where $\Z(\cdot)\coloneqq Z\cdot Z$ with $Z=\dm{0}-\dm{1}$ being the Pauli $Z$ matrix, and $\Pi_{\dm{1}}(\cdot)\coloneqq \dm{1}\cdot\dm{1}$ being the projection onto the state $\ket{1}$.


\section{Discussion}

We introduced a comprehensive quantitative approach to the study of non-completely-positive linear maps, focusing in particular on the task of approximating and simulating them with valid quantum channels. To this end, we considered several quantifiers which generalise measures employed in the study of quantum resources --- namely, variants of the robustness and base norm measures. We showed that they satisfy very close relations with the diamond norm, and in particular are exactly equal to it for any trace-preserving linear map. Since such trace-preserving maps are the most commonly encountered examples of dynamics beyond physical quantum channels, this allowed us to establish fruitful interrelations between the quantities, and discover new applications of the fundamentally important quantity that is the diamond norm. We developed in particular two operational connections. Firstly, we introduced a method of simulating general linear maps with quantum channels, shifting the difficulty of realising non-quantum dynamics onto the structurally simpler task of implementing linear combinations of quantum states. We showed that our robustness measure exactly quantifies the cost of realising such schemes in terms of the required state-based resources. Secondly, we showed that another variant of the robustness finds use as an exact quantifier of the performance advantage that a general linear map can enable over quantum channels in a class of state discrimination games. We introduced a number of useful bounds and explicitly employed them to demonstrate the computability of the measures for some representative examples. Finally, we showed how our measures can find use in the quantitative characterisation of several practically relevant settings, namely, structural approximations of positive maps, non-Markovianity quantification, and tightly bounding the cost of probabilistic error mitigation.

Although we focused on the application of our framework to Hermiticity-preserving maps, we note that more general linear maps can be treated in a similar way. The simplest way to approach this is to decompose any linear map $\Phi$ into its Hermiticity-preserving and skew-Hermiticity-preserving parts, that is, write $\Phi = \Phi_{\rm H} + i \Phi_{\rm SH}$ where the constituent maps are defined through $J_{\Phi_{\rm H}} \coloneqq \frac12 (J_\Phi + J_\Phi^\dagger)$ and $J_{\Phi_{\rm SH}} \coloneqq \frac{1}{2i} (J_\Phi - J_\Phi^\dagger)$. The maps $\Phi_{\rm H}$ and $\Phi_{\rm SH}$ are then explicitly Hermiticity-preserving, and our arguments can be applied to them directly. A similar approach was employed in~\cite{buscemi_2013-2} to decompose the two-point quantum correlator $\T : \mathbb{L}(A) \to \mathbb{L}(A \otimes A)$, defined as the map satisfying $\Tr [ \T(\rho) (A \otimes B) ] = \Tr [ A \rho B ]$ for all $A,B$. Indeed, one can show that the decompositions constructed in~\cite{buscemi_2013-2} are also optimal for the robustness-based quantities.

We also note that the diamond norm has been applied as a measure of specific properties of quantum channels, such as their ability to detect coherence~\cite{theurer_2019}. Connections between our methods and such approaches could be fruitful to explore.

A major outstanding issue is to understand how the framework of this work can be extended to non-linear maps, which could allow for the characterisation and more efficient approximation of important unphysical dynamics such as quantum cloners. This question was already asked in the earliest works concerned with approximating non-CPTP maps with quantum channels~\cite{horodecki_2003-4}, but it still remains a considerable challenge to devise approaches which could apply to general non-linear transformations.

\begin{acknowledgments}
We acknowledge fruitful discussions with Joonwoo Bae, Francesco Buscemi, Ludovico Lami, Varun Narasimhachar, Jayne Thompson, and Xiao Yuan. This research is supported by the National Research Foundation (NRF), Singapore, under its NRFF Fellow program (Award No. NRF-NRFF2016-02), the National Research Foundation and Agence Nationale de la Recherche
joint Project No. NRF2017-NRFANR004 VanQuTe, the Singapore Ministry of Education Tier 1 Grant RG162/19 (S) and grant No. FQXi-RFP-IPW-1903 from the Foundational Questions Institute and Fetzer Franklin Fund (a donor advised fund of Silicon Valley Community Foundation). B.R.\ is supported by the Presidential Postdoctoral Fellowship from Nanyang Technological University, Singapore. Any opinions, findings and conclusions or recommendations expressed in this material are those of the author(s) and do not reflect the views of National Research Foundation, Singapore.
\end{acknowledgments}


\let\L\LL
\let\l\ll
\bibliographystyle{unsrtnat}
\bibliography{bib_bartosz,bib_ryuji}

\appendix

\section{Dual forms}\label{app:duals}

Here we derive the dual expressions of the measures, as stated in Prop.~\ref{prop:duals}. The derivation follows standard arguments in convex optimisation~\cite{boyd_2004,ponstein_2004} (see also~\cite[App.\ B]{takagi_2019}). Let us explicitly consider the case of the diamond norm. As our starting point, we will take the primal optimisation problem as in Lem.~\ref{lemma:diamond_herm}:
\begin{equation}\begin{aligned}
  \norm{\Phi}{\dia,p} = \min \lset \mu \bar J_\Phi = M_+ - M_-,\; M_{\pm} \geq 0,\; \Tr_B (M_+ + M_-) \leq \mu \id_{A}  \rset.
\end{aligned}\end{equation}
The Lagrangian of this problem is given by
\begin{equation}\begin{aligned}
  L(\mu, M_+, M_-; W, P, Q, R) &= \mu - \< M_+ - M_- - J_\Phi, W \> - \< M_+, P \> - \< M_-, Q \> \\
  &\;\; - \< \mu \id_A - \Tr_B (M_+ + M_-), R \> \\
  &= \mu ( 1 - \Tr R ) + \< M_+, - W - P + R \otimes \id_B \> \\
  &\;\; + \< M_-, W - Q + R \otimes \id_B \> + \< J_\Phi, W \>
\end{aligned}\end{equation}
where $W, P, Q \in \HH(A \otimes B), R \in \HH(A)$ are Lagrange multipliers, and we used that $\< S, \Tr_B T \> = \< S \otimes \id_B, T \>$ holds for any $S \in \HH(A) , T\in\HH(A \otimes B)$. The dual problem is then defined as
\begin{equation}\begin{aligned}
  \norm{\Phi}{\dia,d} \coloneqq& \sup_{\substack{W \in \HH\\P,Q,R \geq 0}} \inf_{\substack{\mu \in \RR\\M_+, M_- \in \HH}} L(\mu, M_+, M_-; W, P, Q, R)\\
  =& \sup_{\substack{W \in \HH\\P,Q,R \geq 0}} \begin{cases} \< J_\Phi, W \> & \text{ if } \Tr R = 1 \text{ and } W + P = R \otimes \id_B \text{ and } W - Q = -R \otimes \id_B \\
                                                              -\infty & \text{ otherwise} \end{cases}\\
  =& \sup \lset \< J_\Phi, W \> \bar W \geq - R \otimes \id,\; W \leq R \otimes \id, \; R \geq 0,\; \Tr R =1 \rset,
\end{aligned}\end{equation}
with the supremum achieved since the feasible set is compact. A strictly feasible solution, that is, a feasible solution for which the inequality constraints are strict, can be constructed by decomposing $J_\Phi = {J_{\Phi}}_+ - {J_{\Phi}}_-$ and defining $M_\pm \coloneqq {J_{\Phi}}_\pm + \ve \id_{A\otimes B}$ with $\mu$ suitably large. By Slater's theorem (see e.g.~\cite{ponstein_2004}), the existence of a strictly feasible solution ensures that $\norm{\Phi}{\dia,p} =\norm{\Phi}{\dia,d}$.

The dual forms of the other measures are obtained in full analogy with the derivation above. The crucial observation is that an optimisation of the form
\begin{equation}\begin{aligned}
  \norm{\Phi}{\bnorm} = \min \lset \lambda_+ + \lambda_- \bar \Phi = \lambda_+ \Lambda_+ - \lambda_- \Lambda_-,\; \Lambda_{\pm} \in \CPTN \rset
\end{aligned}\end{equation}
can be rewritten as
\begin{equation}\begin{aligned}
  \norm{\Phi}{\bnorm} = \min \lset \lambda_+ + \lambda_- \bar J_{\Phi} = M_+ - M_-,\; M_\pm \geq 0,\; \Tr_B M_\pm \leq \lambda_\pm \id_{A} \rset
\end{aligned}\end{equation}
which allows us to follow the same approach.

\end{document}